\documentclass[journal,final,twocolumn,10pt,twoside]{IEEEtranTMBMC}

\normalsize

\usepackage{tikz}
\usetikzlibrary{shapes,arrows}
\usepackage{graphicx}
\usepackage{caption}
\usepackage{amsmath}
\usepackage{float}
\usepackage{cleveref}
\usepackage{afterpage}
\usepackage{placeins}
\usepackage{cases}
\usepackage{multirow}
\usepackage{amssymb}
\usepackage{algorithm}
\usepackage{algpseudocode}
\usepackage{array}
\usepackage[noadjust]{cite} 
\usepackage{dblfloatfix}    
\begin{document}
\bstctlcite{IEEEexample:BSTcontrol} 
%
\title{A Molecular Communication Perspective on Airborne Pathogen Transmission and Reception via Droplets Generated by Coughing and Sneezing}
%
%

\author{Fatih~Gulec, Baris~Atakan  \thanks{This work was supported by the Scientific and Technological Research Council of Turkey (TUBITAK) under Grant 119E041.}
	\thanks{The authors are with the Department
		of Electrical and Electronics Engineering, Izmir Institute of Technology, 35430, Urla, Izmir, Turkey. (email: fatihgulec@iyte.edu.tr; barisatakan@iyte.edu.tr)}}
\maketitle

\begin{abstract}
Infectious diseases spread via pathogens such as viruses and  bacteria. Airborne pathogen transmission via droplets is an important mode for infectious diseases. In this paper, the spreading mechanism of infectious diseases by airborne pathogen transmission between two humans is modeled with a molecular communication perspective. An end-to-end system model which considers the pathogen-laden cough/sneeze droplets as the input and the infection state of the human as the output is proposed. This model uses the gravity, initial velocity and buoyancy for the propagation of droplets and a receiver model which considers the central part of the human face as the reception interface is proposed. Furthermore, the probability of infection for an uninfected human is derived by modeling the number of propagating droplets as a random process. The numerical results reveal that exposure time and sex of the human affect the probability of infection. In addition, the social distance for a horizontal cough should be at least $\mathbf{1.7}$ m and the safe coughing angle of a coughing human to infect less people should be less than -$\mathbf{25^\circ}$.
\end{abstract}

\begin{IEEEkeywords}
Airborne pathogen transmission, molecular communication, expiratory droplet reception, social distance.
\end{IEEEkeywords}

%

\section{Introduction}
\IEEEPARstart{M}{olecular} communication (MC) is an emerging research area which employs chemical signals for information transfer. Originally, MC is proposed as the communication method of the nanomachines mimicking the biological cells in microscale in a nanonetwork \cite{atakan2014molecular}. MC can also be helpful for practical applications in macroscale \cite{farsad2016comprehensive}.

The pioneering experimental study in macroscale MC is about establishing a communication link using an electrical sprayer, an alcohol sensor and alcohol molecules as the transmitter (TX), receiver (RX) and messenger molecules, respectively \cite{farsad2013tabletop}. For this experimental setup, several channel models are proposed in \cite{farsad2014channel, kim2015universal, gulec2020fluid} and its data rate is improved via multiple input multiple output (MIMO) technique  in \cite{koo2016molecular}. Furthermore, it is shown that MC can be used in macroscale environments where there is significant attenuation for electromagnetic wave-based communication \cite{qiu2014molecular, guo2015molecular}. In \cite{farsad2017novel}, an experimental platform consisting of pumps, pipes and a pH meter is proposed. This platform encodes the information symbols according to the pH level, as also used in a macroscale fluidic platform explained in \cite{khaloopour2019experimental}. Magnetic nanoparticles which are sensed by a susceptometer (as a RX) are employed to encode information symbols in \cite{unterweger2018experimental}. In \cite{mcguiness2018parameter}, a platform in which a mass spectrometer (as a RX) and an odor generator (as a TX) communicate is developed. \cite{ozmen2018high} proposes a platform with a chemical vapor transmitter and photoionization detectors. Moreover, laser induced florescence technique is implemented as an experimental platform for macroscale MC \cite{abbaszadeh2019mutual}. 

All of the aforementioned macroscale platforms are generally proposed to develop more efficient MC methods. However, macroscale MC can also be employed to solve practical problems such as finding the distance to a source or the location of the source. For the test-bed given in \cite{farsad2013tabletop}, statistical distance estimation methods are proposed in \cite{gulec2020distance} by using the features extracted from molecular signals with the proposed feature extraction algorithm. For a long range underwater scenario it is theoretically shown that a molecular TX can be localized with a mobile search robot \cite{qiu2015long}. The localization of a passive molecular TX is proposed by using a clustered localization algorithm for the proposed sensor network-based platform \cite{gulec2020localization}. In \cite{gulec2020fluid2}, the distance is estimated by considering liquid droplets as information carriers. The experimental study in \cite{hamidovic2019information} shows that encoding information in droplets is possible.

Hence, MC can be employed as a tool to model the biological phenomena which consider droplets such as the transmission and reception of pathogens (viruses, bacteria, etc.) which cause contagious diseases via droplets. This concept is first proposed in \cite{khalid2019communication} which consider the infectious human as a blind TX emitting pathogen-laden droplets and sensors as the RX for outdoor environments. This study is improved in \cite{khalid2020modeling} by taking silicon nanowire field effect transistor-based biosensors into account to model pathogen detection for airborne pathogen transmission. 

Airborne transmission and self-inoculation (direct contact) are two modes of infectious disease transmission via pathogens \cite{bourouiba2014violent}. Droplets can be classified as large droplets and droplet nuclei (aerosols) which have sizes of larger and smaller than $10$ $\mu$m, respectively \cite{ai2018airborne}. All expiratory activities such as coughing, sneezing, breathing and speaking can generate large droplets and aerosols. While large droplets can be effective in short-range, aerosols can spread pathogens to longer distances due to their interactions with air. Airborne transmission via droplets is a significant infection mechanism for pathogens such as influenza virus \cite{killingley2013routes}, severe acute respiratory syndrome (SARS) virus \cite{peiris2003severe} and new SARS coronavirus-2 (SARS-CoV-2) which causes coronavirus disease 2019 (COVID-19) \cite{prather2020reducing}. By the time this paper is written, the global pandemic of COVID-19  still continues and there is no cure for COVID-19. It is essential to emphasize that this study is related with the COVID-19 outbreak, since one of the main mechanisms of this disease is airborne transmission. In this paper, we investigate airborne pathogen transmission and reception mechanisms between humans with a MC perspective for indoor environments. This MC perspective leads us to use and adapt the well-known communication engineering techniques to model the spread of the infectious diseases between humans. The infectious human which emit a cloud consisting of pathogen-laden droplets and air by coughing/sneezing is considered as the TX and the uninfected human is defined as the RX unlike the studies in \cite{khalid2019communication,khalid2020modeling} where the RX is a biosensor. Furthermore, the effect of gravity and buoyancy is taken into account for the indoor propagation of the cloud, which is not considered in \cite{khalid2020modeling}. The propagation of the cloud is modeled by modifying the deterministic model in \cite{bourouiba2014violent} in a probabilistic way. The cloud  travels under the influence of initial velocity, buoyancy and gravity. The number of droplets in this cloud is modeled as a random process. A receiver model which takes the central part of human face as the interface with pathogens into account is proposed. The propagation and reception models are employed for the proposed end-to-end system model in order to give the infection state of the RX as the system output. The contributions of this paper can be summarized as follows:
\begin{itemize}
	\item MC perspective is proposed to model the airborne pathogen transmission and reception mechanisms by humans.
	\item The interaction of pathogen-laden droplets and the uninfected human is examined by proposing a RX model. Hence, an end-to-end system model is proposed which combines the channel and receiver models.
	\item A probabilistic approach which enables the derivation of the probability of infection for an uninfected human is employed. 
\end{itemize}
Furthermore, the proposed model is evaluated by numerical results to understand the effects of physical parameters for different indoor scenarios. Our key findings for a coughing TX is given as follows:
\begin{itemize}
	\item Increased exposure time to pathogens increases the probability of infection.
	\item For a horizontal cough, the social distance should be at least  $ 1.7 $ m.
	\item It is safer to cough with an initial angle less than -$ 25^\circ $ to infect less people.
	\item Male humans are more susceptible to airborne infection due to their larger faces for some situations. 	
\end{itemize}

The rest of the paper is organized as follows. In Section \ref{EESM}, the proposed end-to-end system model is presented. Section \ref{POI} provides the derivation of the probability of infection. Numerical results are given in Section \ref{NR} and the paper is concluded in Section \ref{Conc}.
\section{End-to-End System Model} \label{EESM}
\begin{figure}[h] 
	\centering
	\scalebox{0.44}{\includegraphics{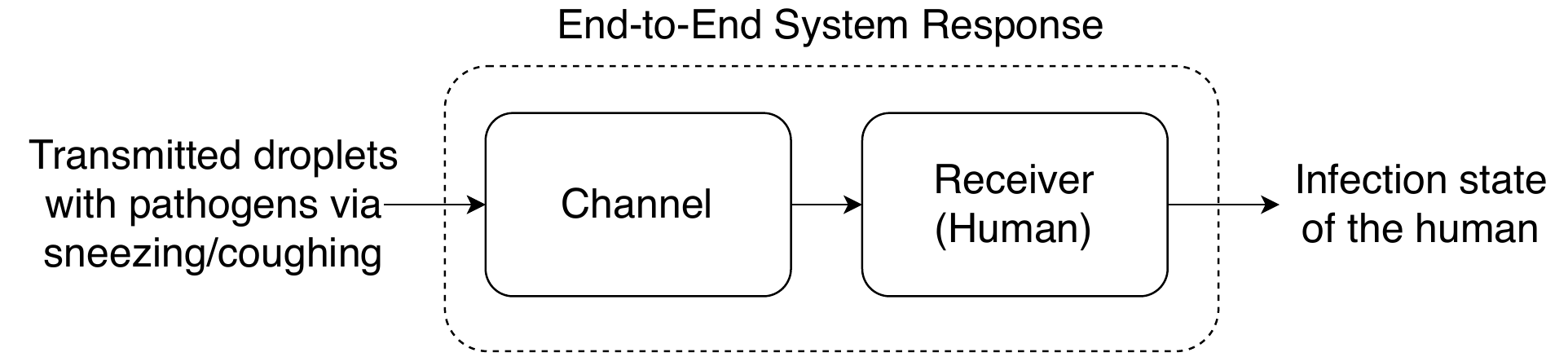}}
	\caption{Block diagram of the end-to-end system model.}
	\label{End_to_end}
	\vspace{-0.5cm}
\end{figure}

This section provides a detailed explanation of the proposed end-to-end system model for droplet-based MC between two humans via sneezing/coughing in four steps. As given in Fig. \ref{End_to_end}, this model incorporates the airborne pathogen-laden droplet transmission with the reception of these droplets by the human that is considered as the RX. Fig. \ref{End_to_end} also shows that the end-to-end system impulse response is defined as the infection state which is the output of the end-to-end system, since a sneeze/cough can be considered as an impulsive input signal. Transmitted droplets via sneezing/coughing are modeled as a cloud which is a mixture of air and droplets. The propagation of the cloud is defined as a two-phase flow where the first and second phase represent the liquid phase of droplets and the gas phase of the air, respectively \cite{munson2009fundamentals}. As the first step of the end-to-end system model, the trajectory of the cloud is derived.  In this study, the model given in \cite{bourouiba2014violent} for the propagation of the cloud is adopted and modified. In the second step, we derive an  end-to-end system model with a probabilistic approach instead of the deterministic approach in \cite{bourouiba2014violent}. The third step details the RX model which includes signal reconstruction, integration, quantization and detection parts. In the last step, the algorithm for the implementation of the system model is given. 
\begin{figure*}[btp]
	\centering
	\scalebox{0.85}{\includegraphics{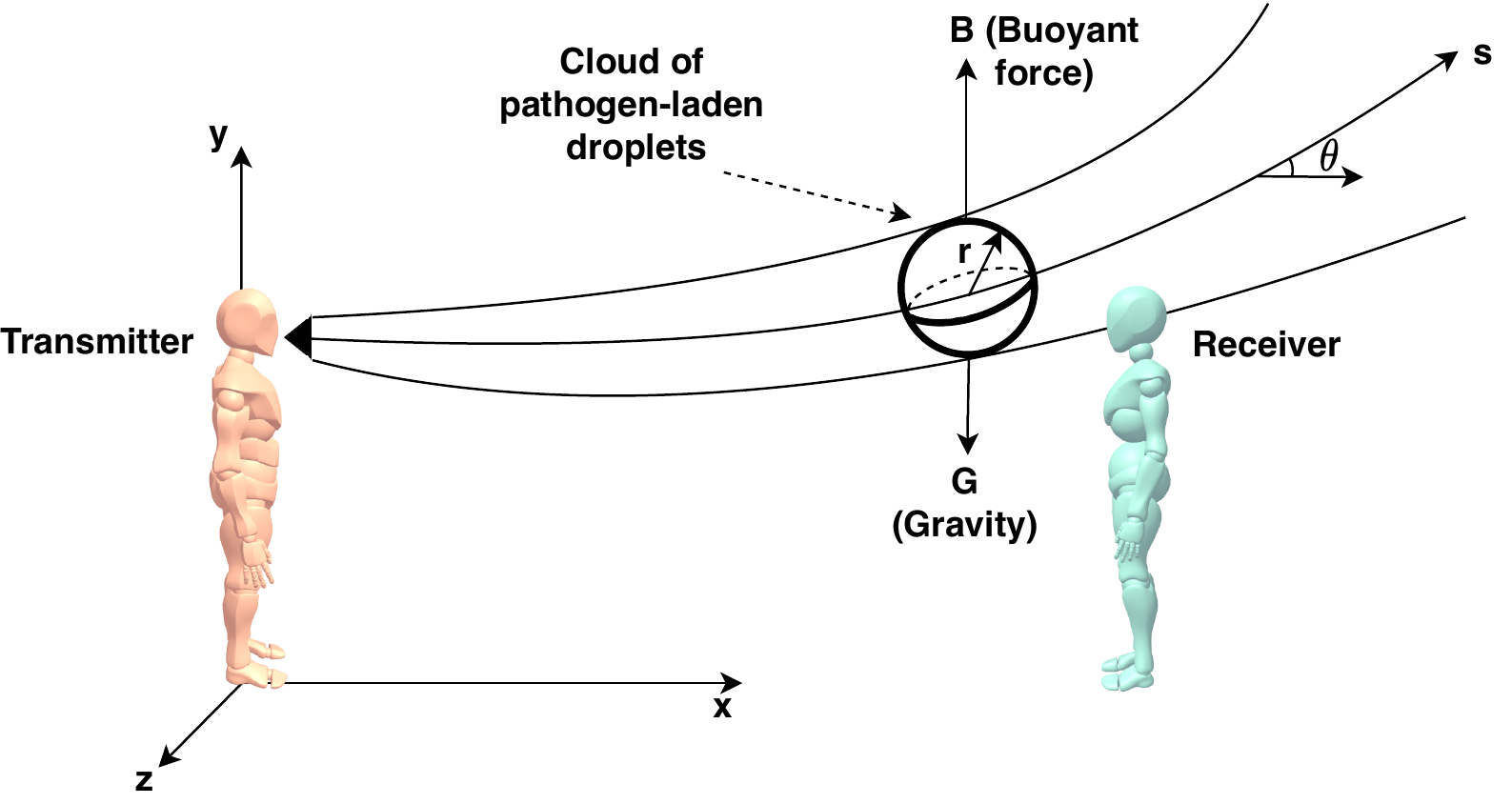}}
	\caption{Trajectory of the cloud between the TX and RX.}
	\label{Trajectory}
\end{figure*}
\subsection{Trajectory of the Transmitted Cloud} \label{TrC}
In our scenario, the TX emits the cloud with an initial velocity on the $x$-axis by sneezing or coughing. Due to the warmer air in the mouth (with density $\rho_f$ at $34^\circ C$) with respect to ambient air (with density $\rho_a$ at $23^\circ C$), where $\rho_f < \rho_a$ \cite{duguid1946size}, the emitted cloud is subject to buoyancy on $y$-axis. As illustrated in Fig. \ref{Trajectory}, buoyancy, gravity and initial velocity of the cloud affect the trajectory of the cloud. Therefore, the trajectory is defined with the curvilinear $ s $-axis and  $\theta$ which shows the angle between the $s$ and $x$ axes. In order to find the position of the cloud in 3-D space, it is essential to derive the time-dependent density of the cloud ($\rho_c(t)$). At the initial state ($t=0$), the initial cloud mass ($m_c(0)$) can be represented by the addition of the initial droplet mass ($m_d(0)$) and initial air mass in the cloud ($m_a(0)$) as given by \cite{munson2009fundamentals}
\begin{align}
m_c(0) &= m_d(0) + m_a(0) \label{m_c}\\
\rho_c(0) V(0) &= \rho_d V_{d}(0)  + \rho_f (V(0) - V_{d}(0)). \label{rho_c1}
\end{align}
where $V(t)$ is the cloud volume, $\rho_d$ is the droplet density, $\rho_f$ is the air density in the mouth. The volume of droplets in the cloud ($ V_{d}(t) $) consists of different sized droplets with diameter $d_k$ and can be defined in terms of the volume fraction of droplets ($\phi_k(t)$) in the cloud as $ V_d(0) = \mathop{\sum}\limits_{k = 1}^{K} \phi_k(0) V(0) $. Here, $K$ is the number of the different droplet sizes, $\phi_k(0) = N_k(0) V_k/V(0)$, $N_k(t)$ and $V_k$ is the number and volume of the spherical droplets of diameter $d_k$, respectively. $N_k(t)$ changes during the propagation of the cloud due to the settling of droplets to the ground as explained in Section \ref{nod}. By substituting $ V_d(0)$ into (\ref{rho_c1}) and solving for $\rho_c(0)$, the initial density of the cloud is derived as 
\begin{equation}
\rho_c(0) = \sum_{k = 1}^{K} (\rho_d - \rho_f) \phi_k(0) + \rho_f. \label{rho_c2}
\end{equation}
As the cloud moves, it entrains the ambient air with density $\rho_a$ and its volume becomes $V(t) = V(0) + V_a(t)$ where $V_a(t)$ is the acquired air volume. Since the initial volume fraction of the air in the mouth with density $\rho_f$ is relatively small in the moving cloud ($10^{-5}$) \cite{duguid1946size}, it is assumed as $V(t) \approx V_a(t)$. After the emission of droplets, $\rho_c(t)$ can be derived via the conservation of mass \cite{munson2009fundamentals}. Hence, the mass of droplets at the time instance $t$ ($m_d(t)$) is equal to the initial mass of droplets ($m_d(0)$) and droplet mass can be expressed by the difference of cloud mass and air mass according to (\ref{m_c}) as given by 
\begin{align}
m_d(t) &= m_d(0) \\ \hspace{-0.3cm}
\left(\hspace{-0.3cm}
\begin{array}{c} 
\text{ Cloud mass } \\ \text{at time $ t $}	
\end{array}
\hspace{-0.3cm}\right) \hspace{-0.1cm} - \hspace{-0.1cm}
\left(\hspace{-0.2cm}
\begin{array}{c}
\text{Air mass} \\ \text{at time $ t $}	
\end{array}
\hspace{-0.2cm}\right) \hspace{-0.1cm} &=\hspace{-0.1cm}
\left(\hspace{-0.2cm}
\begin{array}{c}
\text{Initial} \\ \text{cloud mass}	
\end{array}
\hspace{-0.2cm}\right)\hspace{-0.1cm} - \hspace{-0.1cm}
\left(\hspace{-0.2cm}
\begin{array}{c}
\mbox{Initial} \\ \text{air mass}	
\end{array}
\hspace{-0.2cm}\right) \hspace{-0.15cm}\\
\rho_c(t) V(t) - \rho_a V(t) &= \rho_c(0) V(0) - \rho_f V(0). \label{rho_c3}
\end{align}
Via the substitution of (\ref{rho_c2}) into (\ref{rho_c3}) and some algebraic manipulation, the cloud density is derived as
\begin{equation}
\rho_c(t) = \sum_{k = 1}^{K} (\rho_d - \rho_f) \phi_k(t) + \rho_a.
\label{rho_c4}
\end{equation} 

As illustrated in Fig. \ref{Trajectory}, there are two acting forces on the cloud which stem from the gravity and buoyancy on $y$-axis. Since $\rho_f < \rho_a $, the buoyant force ($B(t)$) affects the movement of the cloud upwards. The net buoyant force acting on the cloud ($F_B(t)$) on $y$-axis is given by the difference of $B(t)$ and the gravitational force ($G(t)$) as given by \cite{munson2009fundamentals}
\begin{align}
F_B(t) &= B(t) - G(t) = V(t) \rho_a g - V(t) \rho_c(t) g \label{F_B1}\\
	&=  V(t)(\rho_a - \rho_c(t)) g \label{F_B2}
\end{align}
where $g$ is the gravitational acceleration.

\begin{figure*}[!tp]
	\normalsize
	\begin{multline}
	\frac{\eta \alpha^3 \rho_a}{4}s(t)^4 + Z s(t) - \left(\frac{F_0 t + I_0 sin(\theta_0)}{2 F_0}\right) \sqrt{F_0^2 t^2 + 2 F_0 I_0 sin(\theta_0) t + I_0^2} \\ -  \left(\frac{I_0^2(sin(\theta_0)^2-1)}{2 F_0}\right) \ln\left(2F_0 \sqrt{F_0^2 t^2 + 2 F_0 I_0 sin(\theta_0) t + I_0^2} + 2F_0^2 t + 2 F_0 I_0 sin(\theta_0)\right) = 0.
	\label{s}
	\end{multline}
		\hrulefill
\end{figure*}

On $ x $-axis, the movement of the cloud is driven by the momentum ($I$) which is defined as the multiplication of the mass and velocity \cite{munson2009fundamentals}. The momentum is not effective on $y$-axis and also there is not any acting force on the cloud for $z-$axis. $I$ is defined on $s$-axis and is decomposed into two components on $x$ ($I_x = |I|  \cos(\theta)$) and $y$ axes ($I_y = |I|\sin(\theta)$). Since the force can be represented as the derivative of the momentum \cite{munson2009fundamentals} and there is not any acting force on the $x$-axis during the propagation, the net force ($F_x$) on  $x$-axis is given by
\begin{equation}
F_x = \frac{d I_x}{dt} = \frac{d |I|\cos(\theta)}{dt} = 0.
\label{Fx}
\end{equation}
Furthermore, the net force on $y$-axis, i.e., $F_B(t)$ is given by
\begin{equation}
F_B(t) = \frac{d I_y}{dt} = \frac{d |I|\sin(\theta)}{dt}.
\label{Fy}
\end{equation}
Since the initial buoyancy is conserved \cite{bourouiba2014violent}, we have $F_B(t) = F_0$ where $F_0$ is the net initial buoyant force. With the initial conditions which are $|I(0)| = I_0$, $\theta(0) = \theta_0$, $I_x(0) = I_0 \cos(\theta_0)$ and  $I_y(0) = I_0 \sin(\theta_0)$, $I_x$ and $I_y$ can be given as the solutions of (\ref{Fx}) and (\ref{Fy}) as 
\begin{equation}
I_x = I_0 \cos(\theta_0), \hspace{1cm} I_y = F_0 t + I_0 \sin(\theta_0).
\label{Ixy}
\end{equation}
Since there is not any acting force on $z$-axis, the momentum can be expressed as $|I| = \sqrt{I_x^2 + I_y^2}$. Due to its definition, the momentum can be written as \cite{munson2009fundamentals}
\begin{equation}
|I| = m_c v_c(t) = \rho_c(t) V(t) v_c(t), \label{I1}
\end{equation}
where we can express the cloud velocity ($v_c(t)$) as the displacement on $s$-axis ($s(t)$) in an infinitesimal time interval, i.e., $v_c(t) = ds(t)/dt$. The cloud volume  is defined as $V(t) = \eta r(t)^3$ where $\eta = 4 \pi/3$ for a spherical cloud. Furthermore, the radius of the cloud ($r(t)$) is linearly related with the distance such that $r(t) = \alpha_e s(t)$ where $\alpha_e$ is the entrainment coefficient and is empirically determined \cite{morton1956turbulent}. Hence, (\ref{I1}) can be rewritten as
\begin{equation}
|I| = \rho_c(t) \eta \alpha_e^3 s(t)^3 \frac{ds(t)}{dt}. \label{I2}
\end{equation}
When $|I| = \sqrt{I_x^2 + I_y^2}$ is incorporated into (\ref{I2}), we have
\begin{equation}
\frac{ds(t)}{dt} = \dfrac{\sqrt{I_x^2 + I_y^2}}{\rho_c(t) \eta \alpha_e^3 s(t)^3}. \label{I3}
\end{equation}
Here, (\ref{rho_c4}) and (\ref{Ixy}) are substituted into (\ref{I3}) as given by
\begin{equation}
\frac{ds(t)}{dt} = \frac{\sqrt{F_0^2 t^2 + 2 F_0 I_0 sin(\theta_0) t + I_0^2}}{\left(\mathop{\sum}\limits_{k = 1}^{K} (\rho_d - \rho_f) \phi_k(t) + \rho_a \right) \eta \alpha_e^3 s(t)^3}. \label{ds}
\end{equation}
Remembering that $V(t) = \eta \alpha_e^3 s(t)^3$ and $\phi_k(t) = \frac{N_k(t) V_k}{V(t)}$, the denominator part of (\ref{ds}) is simplified as
\begin{equation}
\frac{ds}{dt} = \frac{\sqrt{F_0^2 t^2 + 2 F_0 I_0 sin(\theta_0) t + I_0^2}}{Z + \rho_a \eta \alpha_e^3 s^3}, \label{ds2}
\end{equation}
where $Z = \mathop{\sum}\limits_{k = 1}^{K} (\rho_d - \rho_f) V_k N_k(t)$.

For convenience, $\theta_0$ is chosen as $0$ in \cite{bourouiba2014violent}. However, $ \theta_0 \neq 0 $ is also considered in order to observe the effect of the initial cough/sneeze angle in our study. In addition to this, the initial conditions which are $s(0) = 0$ and $t(0) = 0$ are taken into account in the integration of (\ref{ds2}) to obtain the quartic equation as given by (\ref{s}). Since the discriminant of the quartic equation (\ref{s}) is less than zero for physically meaningful parameter values, two of the roots are complex and one of the roots is a real and negative number. Therefore, there is only one possible positive real root which is used as the solution. However, this solution is a very long expression to write in this paper and (\ref{s}) is solved numerically as detailed later in this section. 

In order to determine the trajectory of the cloud, $\theta$ needs to be derived. Via the substitution of $I_x = |I|  \cos(\theta)$ and $I_y = |I|  \sin(\theta)$ into (\ref{Ixy}), two expressions are obtained. Then, by solving these two expressions for $|I|$ and equating them gives the equation below
\begin{equation}
|I| = \frac{I_0 \cos(\theta_0)}{\cos(\theta)} = \frac{F_0 t + I_0 \sin(\theta_0)}{\sin(\theta)}.
\end{equation}
which can be solved for $\theta$ as given by
\begin{equation}
\theta = \tan^{-1} \left(\frac{F_0 t }{I_0 \cos(\theta_0)} + \tan(\theta_0) \right).
\label{theta}
\end{equation}

\setcounter{figure}{3}
\begin{figure*}[!b]
	\centering
	\scalebox{0.5}{\includegraphics{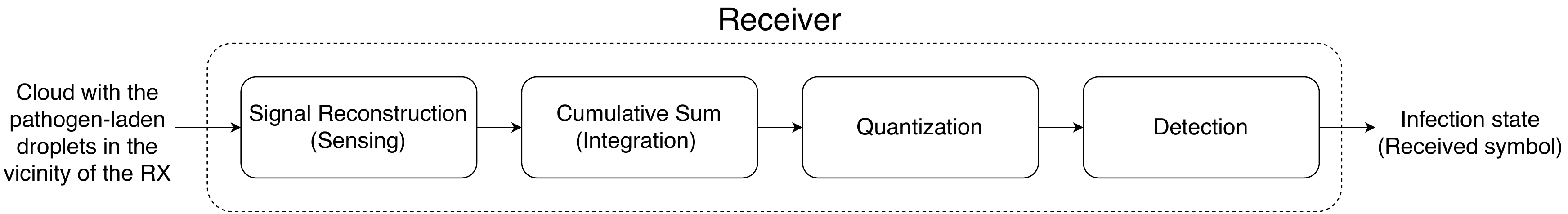}}
	\caption{Block diagram of the receiver model.}
	\label{RX_model}
\end{figure*}
\subsection{Number of Propagating Droplets in the Cloud} \label{nod}
After the emission of droplets, some of the droplets settle to the ground due to gravity and their interaction with the air \cite{gulec2020distance, de2017investigation}. Therefore, the number of droplets decreases during the propagation due to the settling of droplets. In addition, since the movement of each droplet in the cloud is independent of each other at each time instance, the number of droplets in the cloud $(N(t) = \mathop{\sum}\limits_{k = 1}^{K} N_k(t))$ can be modeled as a Poisson process with an intensity function $\lambda(t)$. Due to the large number of emitted droplets, $N(t)$ can be approximated as a Gaussian random process with the same mean and variance, i.e., $N(t) \sim \mathcal{N}(\lambda(t), \lambda(t))$ \cite{papoulis2002probability}. Here, $\lambda(t)$ which is the mean number of droplets in the cloud can be derived by using the flow rate of the droplets \cite{bourouiba2014violent}. Moreover, it is assumed that the droplets are homogeneously distributed within the cloud. The flow rate of the droplets ($J$), which is the derivative of the number of droplets and gives the number of droplets flowing through a surface in unit time (number of droplets/s), is defined as \cite{munson2009fundamentals}
\begin{equation}
J = \frac{d \lambda(t)}{dt} = v_c  A  \bar{\rho}_c(t)
\label{J}
\end{equation}
where $ \bar{\rho}_c(t) = \lambda(t)/V(t)$ and $A$ is the cross-sectional area that droplets are flowing through. Since droplets settle through the lower half of the cloud due to the observations in \cite{bourouiba2014violent}, $A$ in (\ref{J}) is substituted with $A(t)/2$ where $A(t)$ is the surface area of the cloud. Hence, (\ref{J}) becomes
\begin{equation}
\frac{d \lambda(t)}{dt} = - v_s \frac{ A(t)}{2} \frac{\lambda(t)}{V(t)},
\label{rate2}
\end{equation}
where $v_s$ is the settling velocity of droplets and $(-)$ sign represents the decrease in the number of droplets due to the settling. For a spherical cloud, when the substitutions $A(t) = 4 \pi r(t)^2$, $V(t) = 4 \pi r(t)^3/3$ and $r(t) = \alpha_e s(t)$ are made into (\ref{rate2}), the rate of change of the mean number of droplets in the cloud is given by
\begin{equation}
\frac{d \lambda(t)}{dt} = \frac{-3 v_s \lambda(t)}{2 r(t)} = \frac{-3 v_s \lambda(t)}{2 \alpha_e s(t)}.
\label{rate}
\end{equation}

Since the solution of (\ref{s}) is very long and makes the solution of (\ref{rate}) very complicated, the trajectory and number of droplets can be found numerically for each time instance. Furthermore, the number of droplets can be different for each droplet size. Thus, (\ref{rate}) is manipulated to derive the change of the mean number of droplets at each time instance ($\Delta \lambda$) as given by
\begin{equation}
\Delta \lambda = \frac{-3 v_{s_{k,i}} \lambda_{k,i} \Delta t}{2 \alpha s_{k,i}},  
\label{del_la}
\end{equation}
where $\Delta t$ is the time step, the subscripts $i$ and $k$ show the corresponding variables at $t=t_i$ for the droplets of diameter $d_k$. At each time step, $t_i$ is increased by $\Delta t$ and the mean number of droplets of diameter $d_k$ in the cloud is increased via $\lambda_{k,i+1} = \lambda_{k,i} + \Delta \lambda$. For each time step and droplet size, the number of droplets in the cloud becomes a Gaussian random variable as a sample of the Gaussian random process $N(t)$, i.e., $N_{k,i} \sim \mathcal{N}(\lambda_{k,i}, \lambda_{k,i})$.

Settling velocities of droplets during the propagation are defined according to the flow regimes which are Newton's (turbulent) flow, intermediate flow and Stokes (laminar) flow regimes \cite{reuter2005metrics}. These regimes are determined according to Reynolds number ($Re$) which is a dimensionless coefficient showing the flow type of the fluid as defined by \cite{munson2009fundamentals} 
\begin{equation}
Re_{k,i} = \frac{d_{k,i} \rho_a v_{c_{k,i}}}{\mu_a}. \label{Re}
\end{equation}
(\ref{Re}) shows that $Re_{k,i}$ depends on the changing cloud velocity and droplet diameter at each time step.  The settling velocities according to the aforementioned flow regimes are derived in the Appendix. The result of these derivations are given by \cite{reuter2005metrics}
\begin{numcases}
{v_{s_{k,i}} =} 
\frac{g d_{k,i}^2(\rho_d-\rho_a)}{18 \mu_a} &\hspace{-3.6cm}$, Re<2$ \hspace{-3.5cm} \label{v_s1} \\[-7pt]  \hspace{3cm} \textrm{(Stokes flow)} \nonumber \\ 
\frac{g d_{k,i}^{8/5}(\rho_d-\rho_a)}{13.875 \rho_d^{2/5} \mu_a^{3/5}} &\hspace{-3.6cm}$, 2 \leq Re \leq 500$ \label{v_s2} \\[-10pt] \hspace{3cm} \textrm{(Intermediate flow)}  \nonumber \\ 
\frac{3.03 g d_{k,i} (\rho_d-\rho_a)}{\rho_d} &\hspace{-3.6cm}, $500 < Re \leq 2\times 10^5$. \label{v_s3}\\[-7pt] \hspace{3cm} \textrm{(Newton's flow)} \nonumber
\end{numcases}

\subsection{Receiver Model} \label{ROD}
Airborne pathogen transmission via droplets is infectious, since emitted pathogen-laden droplets can be sensed by nose, mouth and eyes \cite{ai2018airborne,peiris2003severe}. Hence, the human face is where the sensing of the infectious pathogens mostly occurs. Even if the pathogens are not directly received via the facial sensory organs, it is possible to become infected by directly touching the face and sensory organs consecutively. With this motivation, the cross-sectional area of the central part of the human face is considered as the RX cross-section as shown in Fig. \ref{RX}. Moreover, a receiver model is proposed for the reception of droplets as shown in Fig. \ref{RX_model}.
\setcounter{figure}{2}
\begin{figure}[ht]
	\centering
	\scalebox{0.7}{\includegraphics{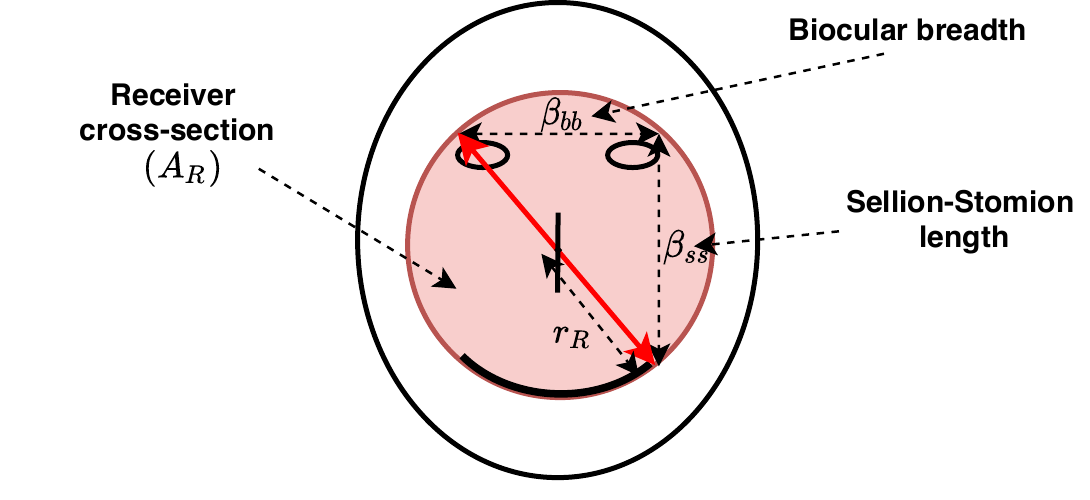}}
	\caption{Receiver cross-section in the human face.}
	\label{RX}
\end{figure}
\setcounter{figure}{4}

As the first step of the reception, the droplets in the vicinity of the RX is sensed by the human, which is defined as the signal reconstruction. Different signal reconstruction models for a sensor in macroscale  and a nanomachine in microscale  are proposed in  \cite{gulec2020fluid} and \cite{atakan2019signal}, respectively. As given in Fig. \ref{RX}, the RX is assumed to be the cross-section of the human face. In order to determine a circular cross-section area ($A_R$) for the RX by encompassing the eyes, mouth and nose, a right-angled triangle whose sides are biocular (biectocanthus) breadth ($\beta_{bb}$), Sellion-Stomion length ($\beta_{ss}$) and the diameter of the receiver cross-section ($2r_R$) as the hypotenuse side is formed as depicted in Fig. \ref{RX}. Here, $\beta_{bb}$  is the length of the line connecting the outer end points of the left and right eyes (eyelid junctions) and $\beta_{ss}$ is the vertical distance between the eye and mouth \cite{young1993head}. Hence, the radius of the cross-sectional area of the RX ($r_R$) is given as $r_R = (\sqrt{\beta_{bb}^2+\beta_{ss}^2})/2$.

\begin{figure}[!bhp]
	\centering
	\scalebox{0.4}{\includegraphics{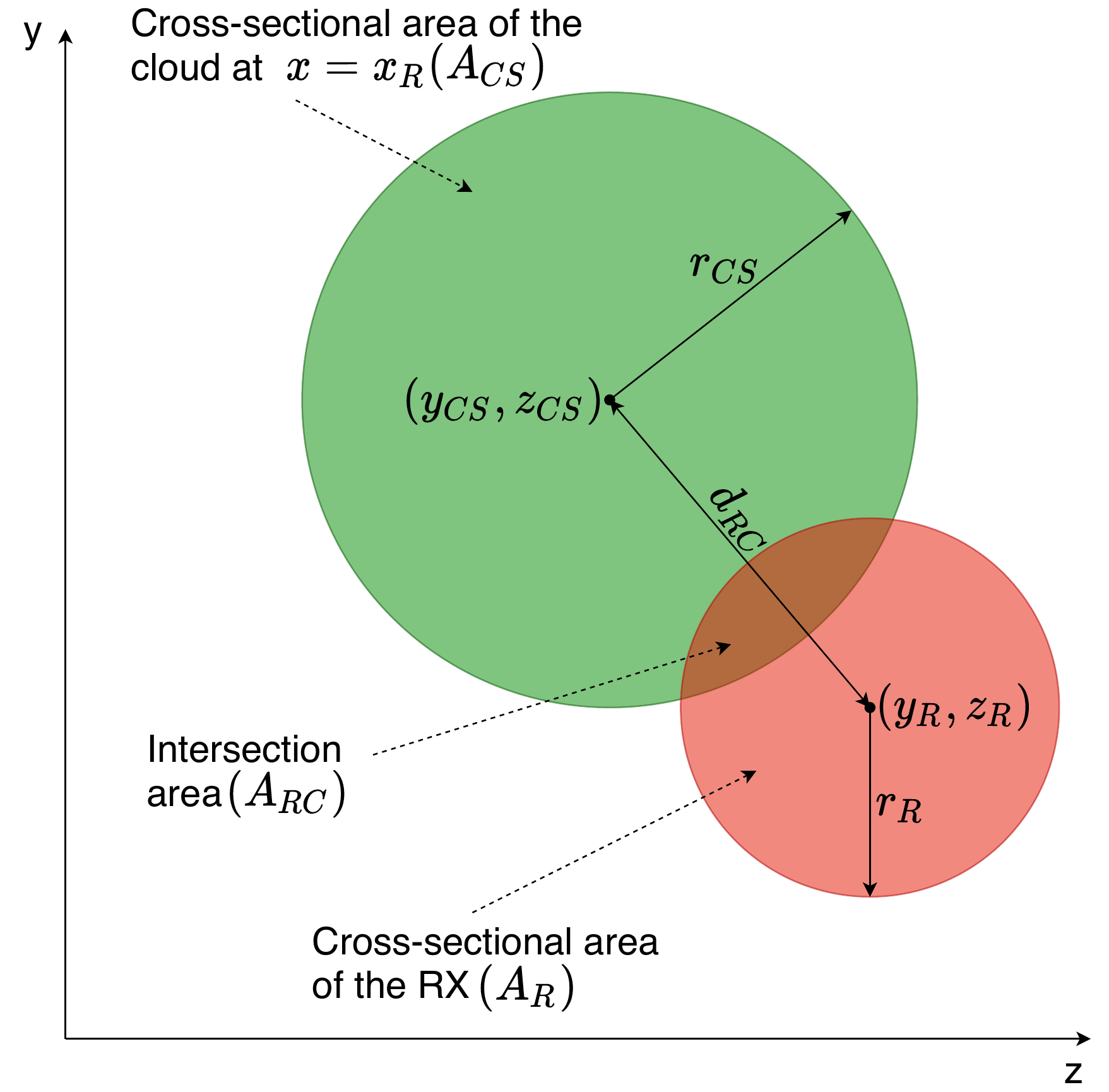}}
	\caption{Intersection of the RX and the cloud cross-section.} 
	\label{IS}
\end{figure}

When the cloud is transmitted via sneezing/coughing, there are three cases for the interaction of the RX and the cloud of droplets with diameter $d_k$ whose centers are at the positions for the $i^{th}$ time step ($x_R$,$y_R$,$z_R$) and ($x_{k,i}$,$y_{k,i}$,$z_{k,i}$), respectively. The relation between the 3-D Cartesian coordinates and curvilinear $s$-axis is detailed in the next subsection. In the first case, the cloud and RX do not coincide and there is no reception. The other two cases include the reception of droplets. As illustrated in Fig. \ref{IS}, when the cloud and RX coincide, the reception of droplets is related with the cross-sectional area of the cloud at $x_{k,i} = x_R$ ($A_{CS_{k,i}}$), $A_R$ and their intersection area ($A_{RC_{k,i}}$). The second case occurs when the intersection area is less than or equal to the cross-sectional area of the RX, i.e., $A_{RC_{k,i}} < A_R$. In the last case, the cloud encompasses the RX, i.e., $A_{CS_{k,i}} \geq A_R=A_{RC_{k,i}}$.

For the case shown in Fig. \ref{IS}, $ A_{RC_{k,i}} $ can be derived by calculating the intersection area of two circles as given by (\ref{A_RC}) \cite{weisstein2003circle}
\begin{figure*}[!tp]
	\normalsize
	\begin{multline}
	A_{RC_{k,i}} = r_R^2 cos^{-1}\left( \frac{d_{RC_{k,i}}^2+r_R^2 - r_{k,i}^2}{2 d_{RC_{k,i}} r_R} \right) + r_{k,i}^2 cos^{-1}\left( \frac{d_{RC_{k,i}}^2+r_{k,i}^2 - r_R^2}{2 d_{RC_{k,i}} r_{k,i}} \right) \\ - \frac{1}{2} \sqrt{(-d_{RC_{k,i}}+r_R+r_{k,i})(d_{RC_{k,i}}+r_R-r_{k,i})(d_{RC_{k,i}}-r_R+r_{k,i})(d_{RC_{k,i}}+r_R-r_{k,i})}.
	\label{A_RC}
	\end{multline}
	\hrulefill
\end{figure*}
where the distance between the centers of $ A_{RC_{k,i}} $ and $A_R$ ($d_{RC_{k,i}}$) is defined as given below.
\begin{equation}
d_{RC_{k,i}} = \sqrt{(y_R - y_{k,i})^2 + (z_R-z_{k,i})^2}.
\label{d_RC}
\end{equation}

The received number of droplets can be derived by multiplying the time step ($\Delta t$) with the flow rate of the droplets ($J$ in number of droplets/s) at each time step. Since the RX senses the droplets proportional to $ A_{RC_{k,i}} $, the received signal after the signal reconstruction step for the aforementioned cases at each time step by recalling (\ref{J}) is expressed as
\begin{numcases}
{\hspace{-1cm}\tilde{N}_{R_i} =} 
\sum_{k = 1}^{K}  v_{c_{k,i}} A_{RC_{k,i}} \frac{N_{k,i}}{\eta r_{k,i}^3} \Delta t &\hspace{-0.5cm}, $ A_{RC_{k,i}} < A_R$  \label{NR1} \\
\sum_{k = 1}^{K} v_{c_{k,i}} A_R \frac{N_{k,i}}{\eta r_{k,i}^3} \Delta t &\hspace{-0.5cm}, $ A_{RC_{k,i}} = A_R$ \label{NR2} \\
0 &\hspace{-0.5cm}, otherwise, \label{NR3}
\end{numcases}
where $ A_{RC_{k,i}} = \pi r_{RC_{k,i}}^2$, $ A_R = \pi r_R^2$, $\eta = 4 \pi/3$ and the volume is $\eta r_{k,i}^3$ for the spherical cloud.

As the time elapses, the cumulative exposure to the pathogen-laden droplets at the RX can be modeled by the cumulative sum (integration) of droplets with respect to time subsequent to the signal reconstruction step as given in Fig. \ref{RX_model}. Afterwards, the signal is quantized, since the rate of the change in the mean number of droplets may not be an integer. The number of droplets after the cumulative sum is rounded to the nearest integer in the quantization step. The received signal after the quantization step is given by
\begin{numcases}
{\hspace{-0.4cm}N_{R_i} \hspace{-0.1cm} = \hspace{-0.1cm}} 
\sum_{k = 1}^{K} \bar{N}_{R_{k,i}} &\hspace{-0.6cm}, $ A_{RC_{k,i}}\hspace{-0.1cm} < \hspace{-0.1cm} A_R$  \label{NR11} \\
\sum_{k = 1}^{K}  \bar{N}_{R_{k,i}} &\hspace{-0.6cm}, $ A_{RC_{k,i}}\hspace{-0.1cm} = \hspace{-0.1cm} A_R$ \label{NR22} \\
0 &\hspace{-0.8cm}, otherwise. \label{NR33}
\end{numcases}
Here, the mean received number of droplets after the quantization step for the $k^{th}$ droplet diameter at the $i^{th}$ time step ($ \bar{N}_{R_{k,i}} $) is defined as
\begin{numcases}
{\hspace{-0.4cm}\bar{N}_{R_{k,i}} \hspace{-0.1cm} = \hspace{-0.2cm}} \hspace{-0.2cm}
\left \lfloor \hspace{-0.1cm} \frac{v_{c_{k,i}} A_{RC_{k,i}} \Delta t}{\eta r_{k,i}^3} \hspace{-0.1cm} \sum_{m = 0}^{i} N_{k,i-m} + \frac{1}{2} \hspace{-0.1cm} \right \rfloor &\hspace{-0.8cm}, $ A_{RC_{k,i}}\hspace{-0.1cm} < \hspace{-0.1cm} A_R$  \label{LR11} \\
\hspace{-0.2cm}\left \lfloor \hspace{-0.1cm} \frac{v_{c_{k,i}} A_R \Delta t}{\eta r_{k,i}^3} \sum_{m = 0}^{i} N_{k,i-m} + \frac{1}{2} \hspace{-0.1cm} \right \rfloor &\hspace{-0.8cm}, $ A_{RC_{k,i}}\hspace{-0.1cm} = \hspace{-0.1cm} A_R$ \label{LR22} 
\end{numcases}
where $ \lfloor . \rfloor $ shows the floor function which maps a variable to the integer less than or equal to this variable. The addition with $\frac{1}{2}$ within the floor function in (\ref{LR11})-(\ref{LR22}) provides the quantization by rounding the number of received droplets to the nearest integer. 

Subsequent to quantization, the infection state of the human needs to be determined as the output of the system as shown by Figs. \ref{End_to_end} and \ref{RX_model}. Therefore, the detection is essential according to a threshold value ($ \gamma $) as the last step of the reception. Physically, $\gamma$ corresponds to the quantity of pathogen-laden droplets that suffice to make a human infected. Furthermore, $\gamma$ depends on the immune system of a human. Thus, detection via the threshold $\gamma$ enables to quantify the strength of the human immune system and  to handle the determination of the infection state as a detection problem. Hence, the infection state (or the received symbol) can be expressed by binary hypothesis testing which is given as
\begin{equation}
N_{R_i} \underset{H_0}{\overset{H_1}{\gtrless}}  \gamma,
\label{Det}
\end{equation}
where the hypotheses $H_0$ and $H_1$ are defined as the situations of no infection as the received symbol $0$ and infection as the received symbol $1$, respectively. Here, the received symbol sequence with $M$ samples is represented as $ \mathbf{h} = [h_0, h_1,...,h_M]$ which also gives the end-to-end system response.
\subsection{Algorithm of the End-to-End System Model}
In this part, the way to obtain the output of the end-to-end system model by employing the derivations made up to here is clarified in the proposed Algorithm \ref{Alg1}. Before the procedure starts,  the initial parameters such as the number of droplet sizes ($K$), total simulation time ($t_s$), initial number of droplets for each droplet size ($N_{(1,..K),0}$), initial momentum ($I_0$), initial net buoyant force ($F_0$), initial velocity of the cloud ($v_{c_0}$), RX dimensions ($\beta_{bb}$, $\beta_{ss}$) and initial positions of the TX and RX are taken as the input. In the algorithm, the propagation, changing number of droplets in the cloud and their interaction with the RX are handled separately at each time step for each droplet size $d_{k}$. As the first step of the algorithm, the trajectory of the spherical cloud is calculated for each time step, i.e., the distance on the $s$-axis and $\theta$ values for $t=t_i$ are calculated by (\ref{s}) and (\ref{theta}). Then, the updated position of the cloud on the $s$-axis is utilized to find its step length ($\Delta s$). As shown in Fig. \ref{Trajectory}, $\Delta s$ can be employed to express the step lengths on $x$ ($ \Delta x $) and $y$ axes ($\Delta y$) as given by
\begin{equation}
\Delta x = \Delta s \hspace{0.05cm} \cos(\theta_{k,i}), \hspace{1cm} \Delta y = \Delta s \hspace{0.05cm} \sin(\theta_{k,i}),
\end{equation}
where $\Delta x$ and $\Delta y$ values are used to update the cloud position on the corresponding axis. Since there is not any acting force on $z$-axis, the center of the cloud maintains its position on this axis. However, the cloud expands on $x$, $y$ and $z$ axes linearly due to the relation $r_{k,i} = \alpha_e s_{k,i}$ as given in Section \ref{TrC}.
\begin{algorithm}[!bth]
	\caption{Algorithm of the End-to-End System Model}
	\label{Alg1} 
	\begin{algorithmic}[1]
		\State \textbf{Input:} $K$, $ t_s $, $\Delta t$, $v_{c_0}$, $I_0$, $F_0$, $\rho_a$, $\rho_d$, $\rho_f$, $\mu_a$, $\alpha_e$, $g$, $d_{(1,...,K)}$, $N_{(1,...,K),0}$, $\theta_0$, $\beta_{bb}$, $\beta_{ss}$, initial positions of the TX and RX, 
		\State $t = 0:\Delta t:t_s $
		\For{$ k = 1:1:K $}
			\For{ $ i = 1:1:\text{length}(t) $}
				\\ \textbf{\Comment{Step 1: Trajectory}} \hbox to 0.235\textwidth{}
				\State Calculate $s_{k,i}$ by the real positive root of (\ref{s})
				\State $r_{k,i} = \alpha_e s_{k,i}$
				\State Calculate $\theta_{k,i}$ by (\ref{theta})
				\State $\Delta s = s_{k,i} - s_{k,i-1}$
				\State $\Delta x = \Delta s \cos(\theta_{k,i})$; $\Delta y = \Delta s  \sin(\theta_{k,i})$
				\State $x_{k,i} = x_{k,i-1}+\Delta x$; $y_{k,i} = y_{k,i-1}+\Delta y$
				\\ \textbf{\Comment{Step 2: Number of Droplets in the Cloud}} \hbox to 0.04\textwidth{}
				\State $ v_{c_{k,i}} = \Delta s/ \Delta t $				
				\State $Re_{k,i} = d_{k,i} v_{c_{k,i}} \rho_a/\mu_a$
				\State Calculate $v_{s_{k,i}}$ according to $Re_{k,i}$ by (\ref{v_s1})-(\ref{v_s3})
				\State Calculate $\Delta \lambda$ by (\ref{del_la})
				\State $\lambda_{k,i} = \lambda_{k,i-1} + \Delta \lambda$
				\State Generate $ N_{k,i} \sim \mathcal{N}(\lambda_{k,i}, \lambda_{k,i})$ 
				\\ \textbf{\Comment{Step 3: Reception}} \hbox to 0.239\textwidth{}
				\If{$(x_R-r_{k,i}) < x_{k,i} < (x_R+r_{k,i}) $)}
				\State $r_{CS_{k,i}} = \sqrt{r_{k,i}^2 - (x_R-x_{k,i})^2 }$; 
				\State $A_{CS_{k,i}} = \pi r_{CS_{k,i}}^2$ 
				\State Calculate $d_{RC_{k,i}}$ by (\ref{d_RC})
					\If{$(r_{CS_{k,i}} - r_R) < d_{RC_{k,i}} < (r_{CS_{k,i}} + r_R)$}
						\State Calculate $\bar{N}_{R_{k,i}}$ by (\ref{LR11}) 
					\ElsIf{$d_{RC_{k,i}} < (r_{CS_{k,i}} - r_R)$}
						\State Calculate $\bar{N}_{R_{k,i}}$ by (\ref{LR22})
					\Else
						\State $\bar{N}_{R_{k,i}} = 0$
					\EndIf
					\State $ N_{k,i} = N_{k,i} - \bar{N}_{R_{k,i}} $	
				\EndIf				
			\EndFor
		\EndFor
		\State Calculate $ N_{R_i} $ by (\ref{NR11})-(\ref{NR33})
		\State Determine the infection state ($\mathbf{h}$) by applying (\ref{Det})
	\end{algorithmic} 
\end{algorithm}
\vspace{-0.5cm}

In the second step of Algorithm \ref{Alg1}, the mean number of droplets in the propagating cloud is calculated. To this end, the cloud velocity ($v_{c_{k,i}}$) at the corresponding time step is calculated by the displacement on $s$-axis. Then, settling velocity ($v_{s_{k,i}}$) is determined by (\ref{v_s1})-(\ref{v_s3}) according to $Re_{k,i}$ which is calculated by using $v_{c_{k,i}}$ and droplet size. $v_{s_{k,i}}$ is exploited to calculate the change in the mean number of droplets ($ \Delta \lambda $) and thus, the mean number of droplets is updated according to this change. Here, the flow type of the cloud found by employing the velocity of droplets affects the number of droplets in the cloud.

The third step of Algorithm \ref{Alg1} describes the reception via the interaction of the cloud with the RX. When the cloud comes to a sufficient distance  to interact with the RX, the radius of the cloud's circular cross-section ($r_{CS_{k,i}}$) is determined by the geometrical relation with the the radius of the cloud and the positions of the TX and RX on the $x$-axis as given by
\begin{equation}
r_{CS_{k,i}} = \sqrt{r_{k,i}^2 - (x_R-x_{k,i})^2 },
\end{equation}
which allows us to calculate the circular area of the cloud's cross-section ($A_{CS_{k,i}}$). During the reception, the case that $A_{CS_{k,i}} \leq A_R$ can also be represented in terms of radii of the circles and the distance between them such that $(r_{CS_{k,i}} - r_R) < d_{RC_{k,i}} < (r_{CS_{k,i}} + r_R)$. In addition, the case for the cloud encompassing the RX ($A_{CS_{k,i}} > A_R=A_{RC_{k,i}}$) can be expressed as $d_{RC_{k,i}} < (r_{CS_{k,i}} - r_R)$. Using these conditions, the mean number of droplets given in (\ref{NR11})-(\ref{NR33}) is calculated. Then, the detection is made according to the threshold $\gamma$ in order to determine the infection state. Next, the probability of infection is derived by using the end-to-end system model.
\begin{figure*}[!btp]
	\normalsize
	\begin{numcases}
	{\hspace{-0.5cm}f_{N_R}(N_{R_i}) \hspace{-0.1cm} = \hspace{-0.2cm}} \hspace{-0.2cm}
	\frac{1}{\sqrt{2\pi \mathop{\sum}\limits_{k = 1}^{K} \mathop{\sum}\limits_{m = 0}^{i} \left \lfloor  \left(\frac{v_{c_{k,i}} A_{RC_{k,i}} \Delta t}{\eta r_{k,i}^3}\right)^2  \lambda_{k,i-m} \hspace{-0.1cm} + \hspace{-0.1cm} \frac{1}{2} \right \rfloor} } \text{exp}\hspace{-0.1cm}\left(\hspace{-0.1cm}-\frac{\left(\hspace{-0.2cm} N_{R_i} \hspace{-0.2cm} - \hspace{-0.2cm} \mathop{\sum}\limits_{k = 1}^{K} \mathop{\sum}\limits_{m = 0}^{i} \left \lfloor \hspace{-0.1cm} \frac{v_{c_{k,i}} A_{RC_{k,i}} \Delta t}{\eta r_{k,i}^3} \lambda_{k,i-m} \hspace{-0.1cm} + \hspace{-0.1cm} \frac{1}{2} \right \rfloor \hspace{-0.1cm} \right)^2}{2\mathop{\sum}\limits_{k = 1}^{K} \mathop{\sum}\limits_{m = 0}^{i} \left \lfloor  \left(\hspace{-0.1cm} \frac{v_{c_{k,i}} A_{RC_{k,i}} \Delta t}{\eta r_{k,i}^3}\right)^2 \hspace{-0.1cm} \lambda_{k,i-m} \hspace{-0.1cm} + \hspace{-0.1cm} \frac{1}{2} \right \rfloor}\hspace{-0.05cm}\right) &\hspace{-0.8cm}, \hspace{-0.1cm}$ A_{CS_{k,i}} \hspace{-0.1cm} < \hspace{-0.1cm} A_R$  \label{fNR11} \\ \hspace{-0.2cm}	 
	\frac{1}{\sqrt{2\pi \mathop{\sum}\limits_{k = 1}^{K} \mathop{\sum}\limits_{m = 0}^{i} \left \lfloor \left( \frac{A_R}{A_{CS_{k,i}}} \right)^2  \lambda_{k,i-m} + \frac{1}{2} \right \rfloor}} \text{exp}\left(\hspace{-0.1cm}-\frac{ \left( N_{R_i} - \mathop{\sum}\limits_{k = 1}^{K} \mathop{\sum}\limits_{m = 0}^{i} \left \lfloor  \frac{A_R}{A_{CS_{k,i}}}  \lambda_{k,i-m} + \frac{1}{2} \right \rfloor  \right)^2}{2\mathop{\sum}\limits_{k = 1}^{K} \mathop{\sum}\limits_{m = 0}^{i} \left \lfloor \left( \frac{A_R}{A_{CS_{k,i}}} \right)^2 \lambda_{k,i-m} + \frac{1}{2} \right \rfloor}\right) &\hspace{-0.8cm}, \hspace{-0.1cm}$ A_{RC_{k,i}}\hspace{-0.1cm} = \hspace{-0.1cm} A_R$ \label{fNR22} \\
	0 &\hspace{-0.8cm}, otherwise. \label{fNR33}
	\end{numcases}
	\hrulefill
\end{figure*}
\section{Probability of Infection} \label{POI}
The probabilistic approach which is considered in the system model enables the derivation of the probability of infection of a human exposed to a sneeze or cough. To this end, it is essential to derive the probability  density function (pdf) of the received number of droplets before the detection. As given in Section \ref{nod}, the number of droplets in the cloud is a Gaussian random variable  for the droplet diameter of $d_k$ and $i^{th}$ time step, i.e., $N_{k,i} \sim \mathcal{N}(\lambda_{k,i}, \lambda_{k,i})$. Hence, its pdf ($ f_{N}(N_{k,i}) $) is given by
\begin{equation}
f_{N}(N_{k,i}) = \frac{1}{\sqrt{2\pi \lambda_{k,i}}} e^{-\frac{(N_{k,i}-\lambda_{k,i})^2}{2\lambda_{k,i}} }.
\end{equation}

Since the received number of droplets before the detection ($N_{R_i}$) is a function of $ N_{k,i} $ as given in (\ref{NR11})-(\ref{LR22}), its pdf ($f_{N_R}(N_{R_i})$) is given in (\ref{fNR11})-(\ref{fNR33}). The probability of infection corresponds to the situation where $N_{R_i} > \gamma$ as given by
\begin{equation}
P(N_{R_i} > \gamma) = \int_{\gamma}^{\infty} f_{N_R}(u) du.
\label{P1}
\end{equation}
The solution for (\ref{P1}) can be derived for the pdf given in (\ref{fNR11})-(\ref{fNR33}) in terms of $Q$-function ($Q(x) = \frac{1}{\sqrt{2\pi}}\int_{x}^{\infty} \text{exp}\left(\frac{-u^2}{2} \right) du$) as given by
\begin{numcases}
{\hspace{-0.4cm}P(N_{R_i} > \gamma) \hspace{-0.1cm} = \hspace{-0.1cm}} 
Q\left(\frac{\gamma}{\Omega_1} - \Omega_1 \right) &\hspace{-0.8cm}, $ A_{RC_{k,i}}\hspace{-0.1cm} < \hspace{-0.1cm} A_R$  \label{PNR11} \\
Q\left(\frac{\gamma}{\Omega_2} - \Omega_2 \right) &\hspace{-0.8cm}, $ A_{RC_{k,i}}\hspace{-0.1cm} = \hspace{-0.1cm} A_R$ \label{PNR22} \\
0 &\hspace{-0.8cm}, otherwise, \label{PNR33}
\end{numcases}
where $\Omega_1$ and $\Omega_2$ are defined as
\begin{align}
\Omega_1 &= \mathop{\sum}\limits_{k = 1}^{K} \mathop{\sum}\limits_{m = 0}^{i} \left \lfloor  \frac{v_{c_{k,i}} A_{RC_{k,i}} \Delta t}{\eta r_{k,i}^3} \lambda_{k,i-m} + \frac{1}{2} \right \rfloor \\
\Omega_2 &= \mathop{\sum}\limits_{k = 1}^{K} \mathop{\sum}\limits_{m = 0}^{i} \left \lfloor  \frac{v_{c_{k,i}} A_R \Delta t}{\eta r_{k,i}^3} \lambda_{k,i-m} + \frac{1}{2} \right \rfloor. 
\end{align}
The derived probability of infection and the system model given in the previous section can be employed to analyze the dynamics of the pathogen transmission as given with the numerical results in the next section.
\begin{figure*}[!tbp]
	\centering
	\includegraphics[width=0.325\textwidth]{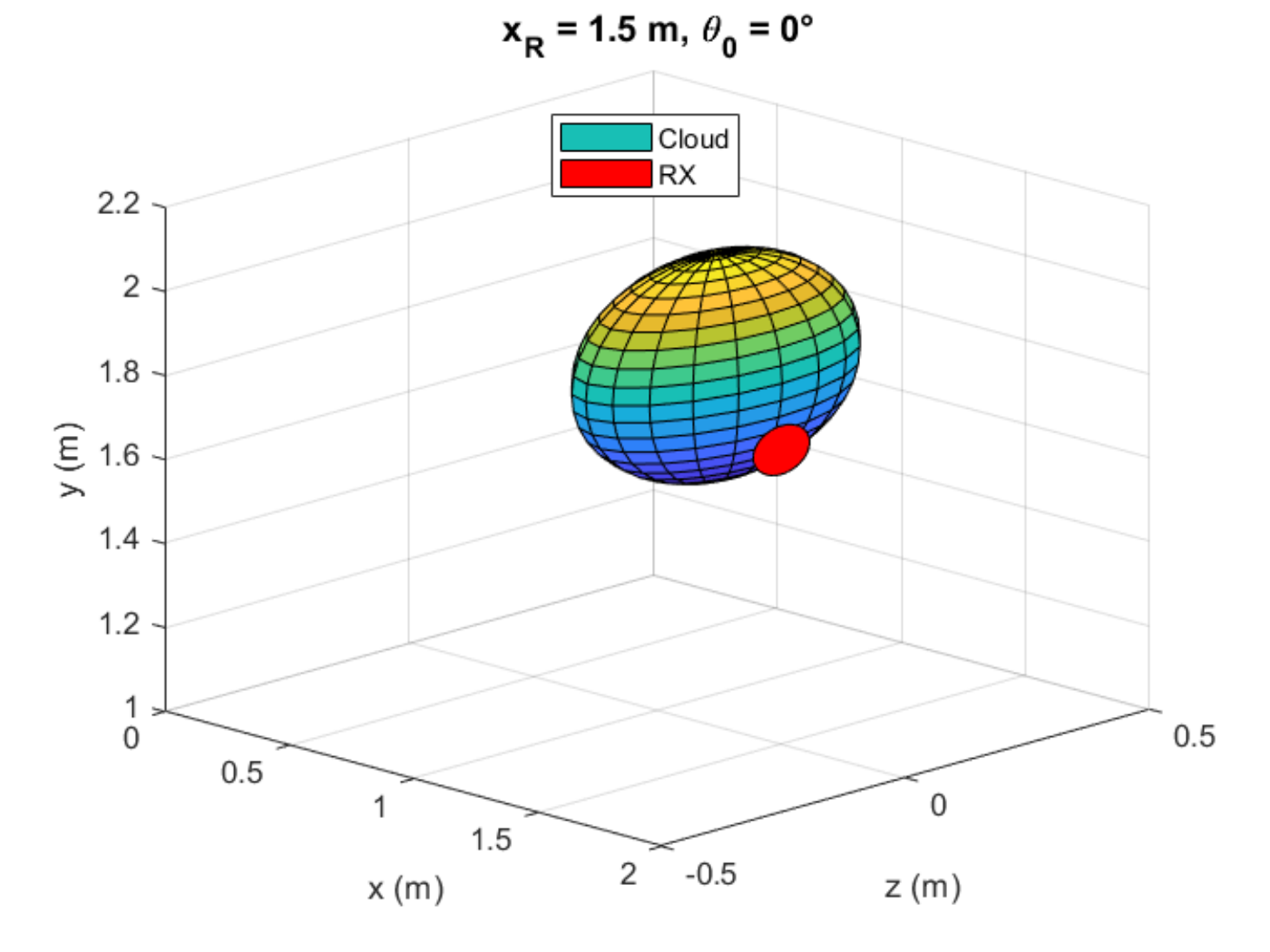}    
	\includegraphics[width=0.325\textwidth]{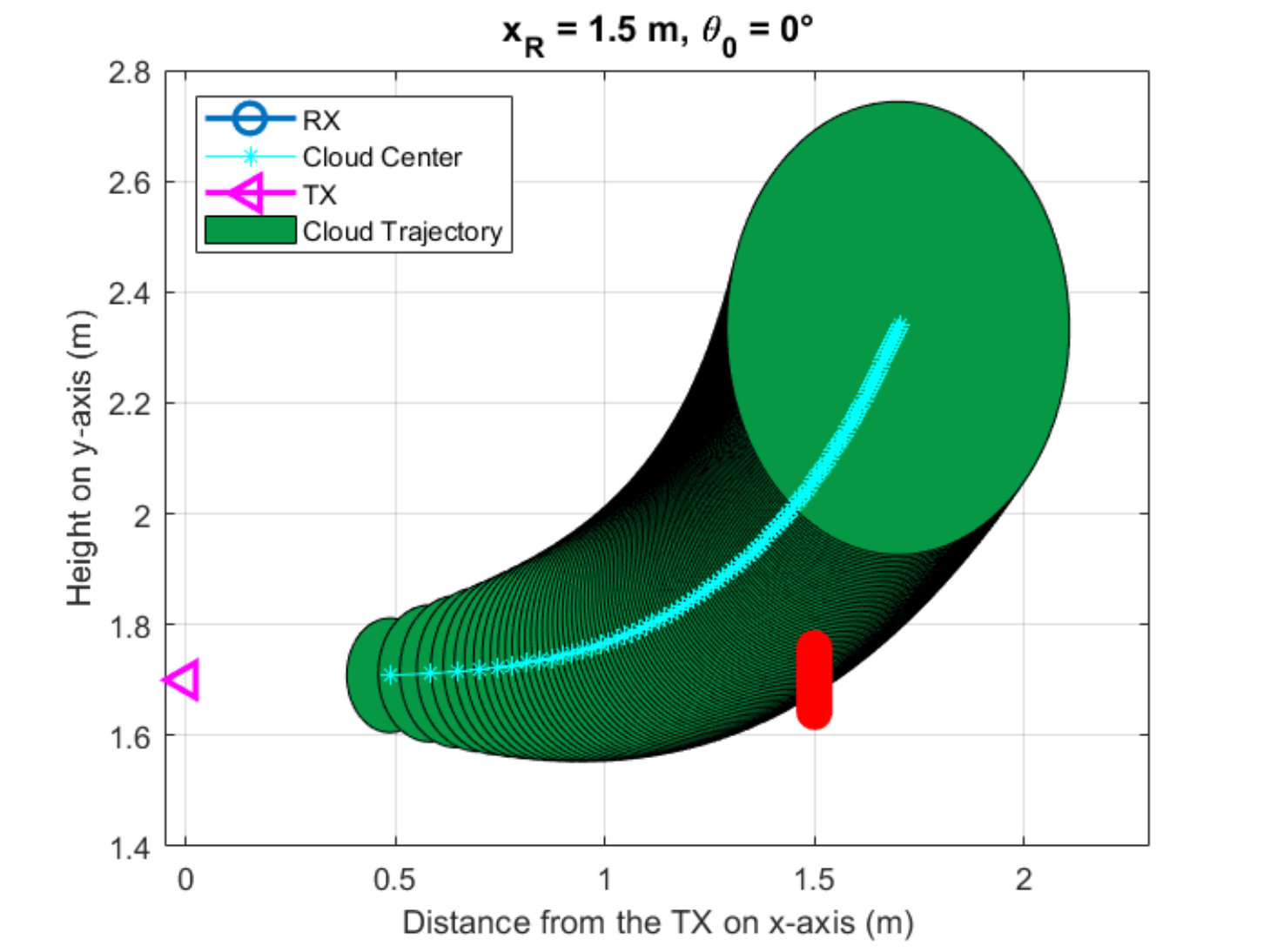}  
	\includegraphics[width=0.325\textwidth]{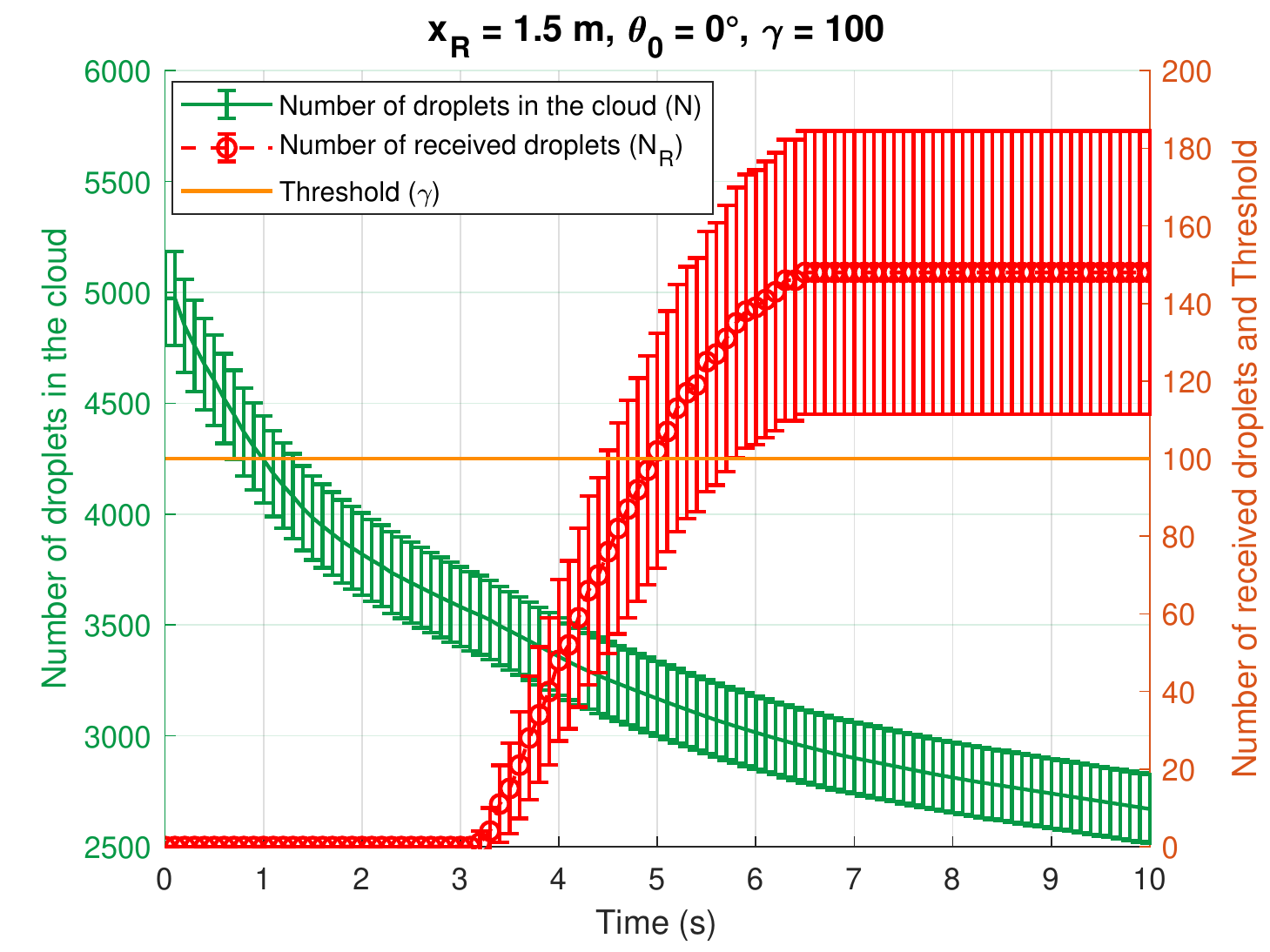} \\
	\scriptsize \hspace{0.1 in} (a) \hspace{2.2 in}  (b) \hspace{2.15 in} (c) \hspace{0.4 in}\\
	\caption{The trajectory of the cloud and its interaction with the RX in (a) 3-D (b) 2-D. (c) The number of droplets.}
	\label{Traj_R}
\end{figure*}

\vspace{-0.3cm}
\section{Numerical Results} \label{NR}
In this section, numerical results using the algorithmic end-to-end system model and derived probability of infection are given. The values of the experimental parameters are given in Table \ref{Sim_parameters}. Except the simulation parameters such as $\Delta t$, $t_s$ and the positions of the TX and RX, this table includes measured values which are obtained by empirical studies \cite{bourouiba2014violent, young1993head, zhu2006study, nicas2005toward, picard2008revised, tang2009schlieren}. Furthermore, the initial number of droplets according to their diameters for a cough is given in Table \ref{Droplet_sizes}. For sneezing, there are not sufficient empirical data in the literature to obtain the parameter values given in Table \ref{Sim_parameters} such as $\alpha_e$ and initial velocity \cite{ai2018airborne}. Therefore, although our proposed model is applicable  to a sneezing scenario, we only consider a coughing scenario for two static humans where one of them is the TX and the other is RX. For the results in Figs. \ref{Traj_R}-\ref{P_inf}, the dimensions of the RX are applied by using the average values of male and female humans. Namely, $ \beta_{bb} $ and $ \beta_{ss} $ are obtained by calculating the average of the female and male values given in Table \ref{Sim_parameters}.
\setcounter{figure}{7}
\begin{figure*}[!b]
	\centering
	\includegraphics[width=0.325\textwidth]{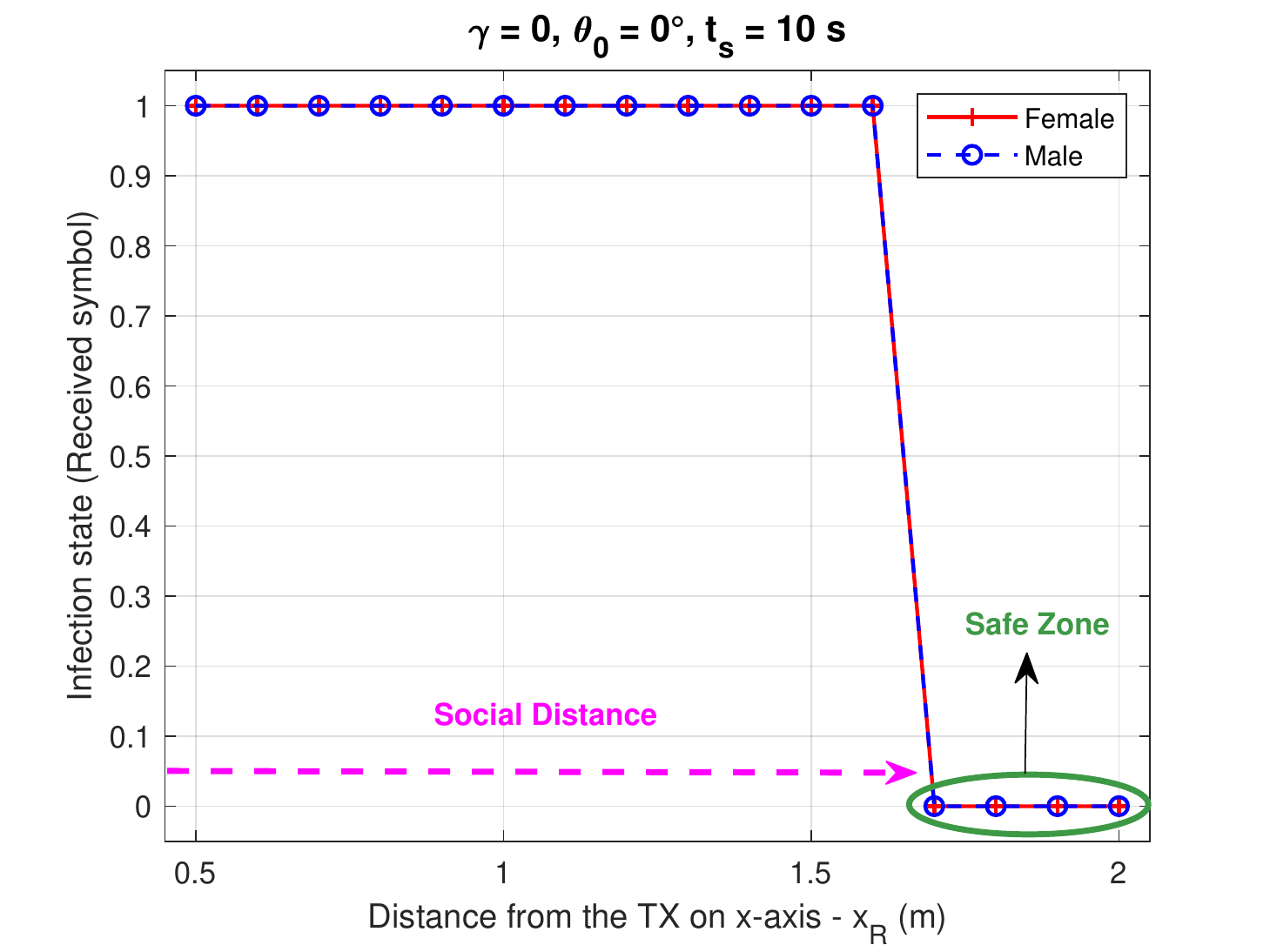}    
	\includegraphics[width=0.325\textwidth]{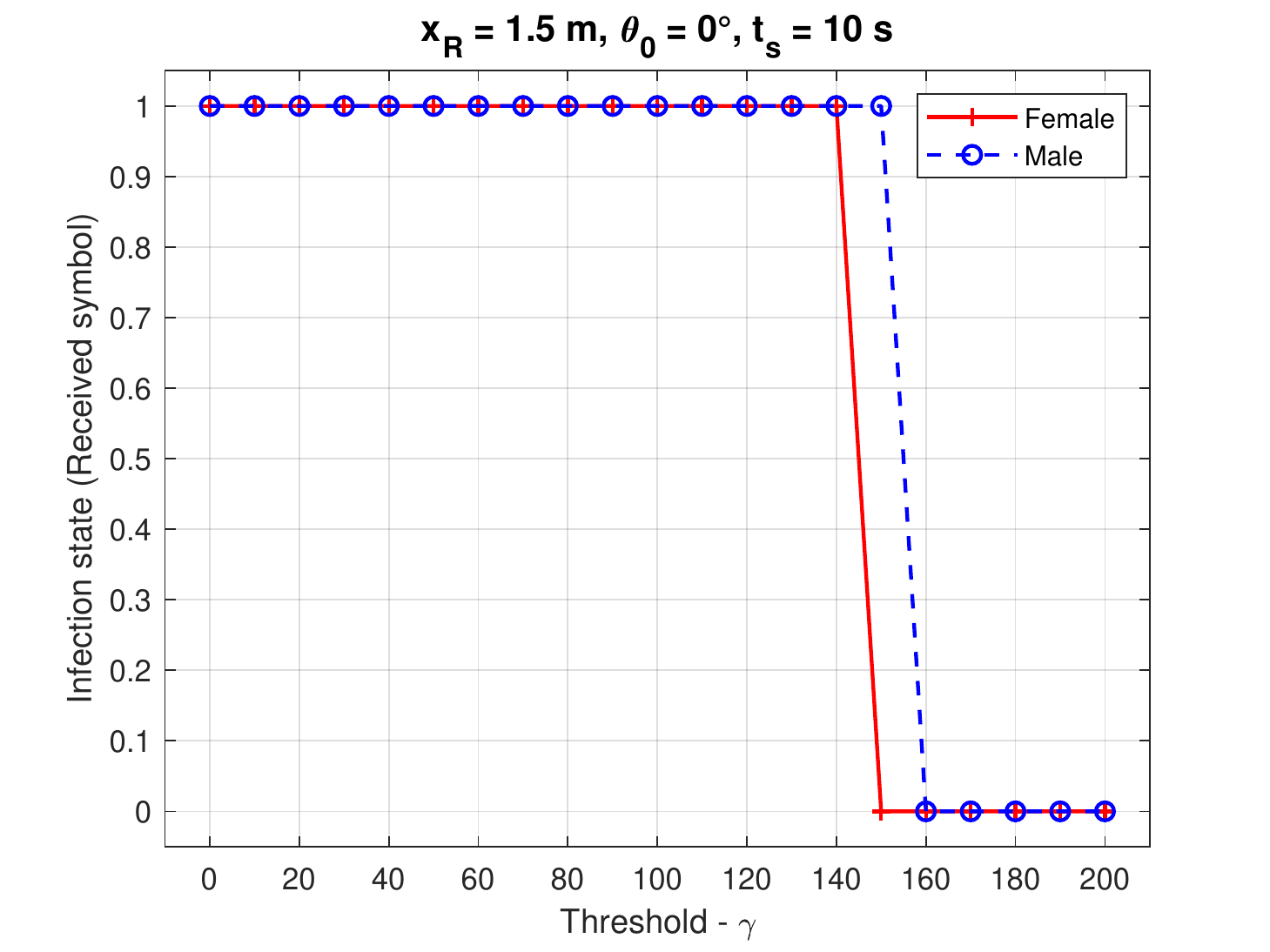}  
	\includegraphics[width=0.325\textwidth]{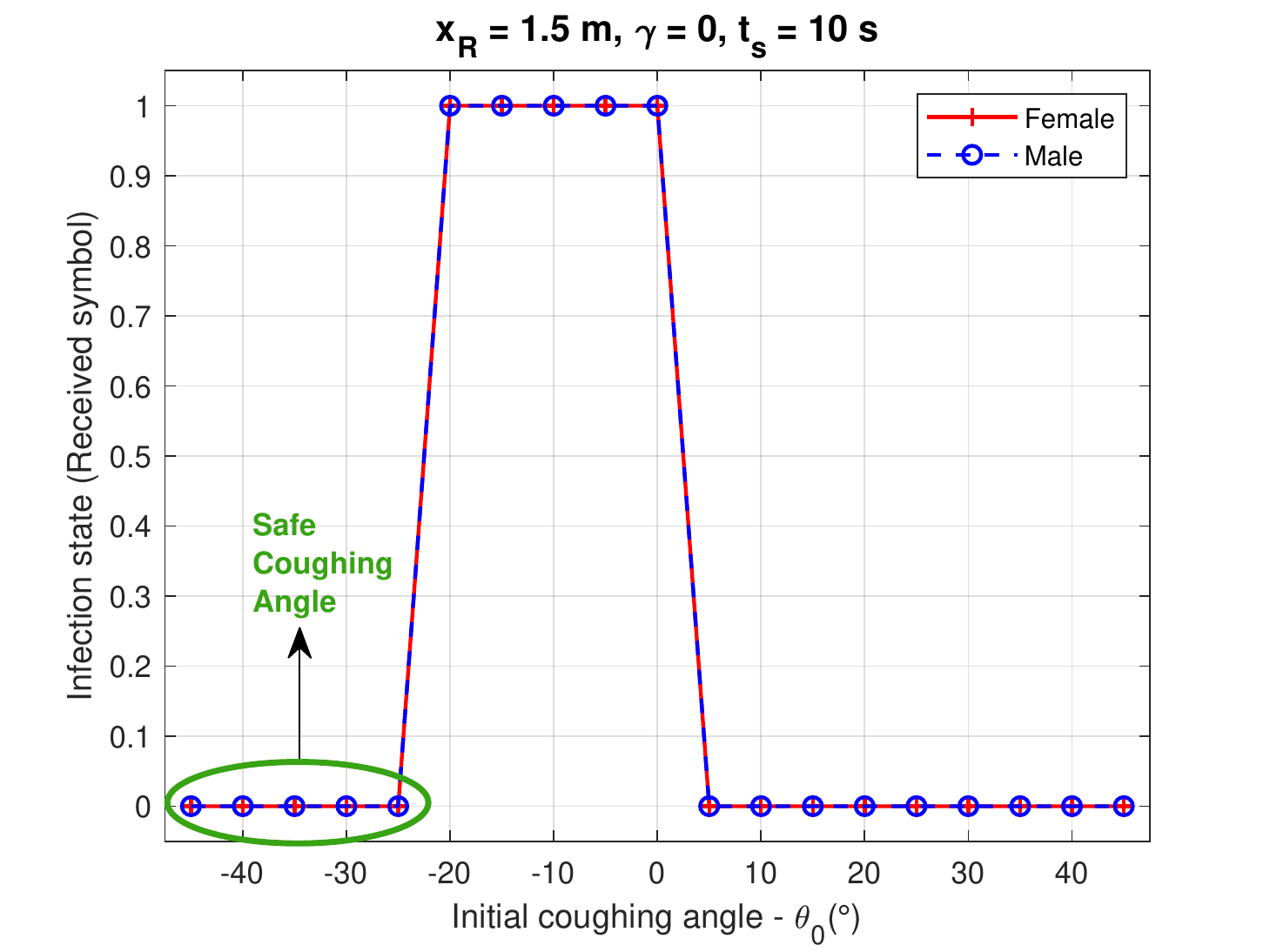} \\
	\scriptsize  (a) \hspace{2.25 in}  (b) \hspace{2.25 in} (c) \hspace{0.4 in}\\
	\caption{Infection state of the RX with respect to (a) $x_R$ (b) $\gamma$ (c) $\theta_0$.}
	\label{Inf_s}
\end{figure*}
\setcounter{figure}{6}
\begin{figure}[!tbph]
	\centering
	\scalebox{0.4}{\includegraphics{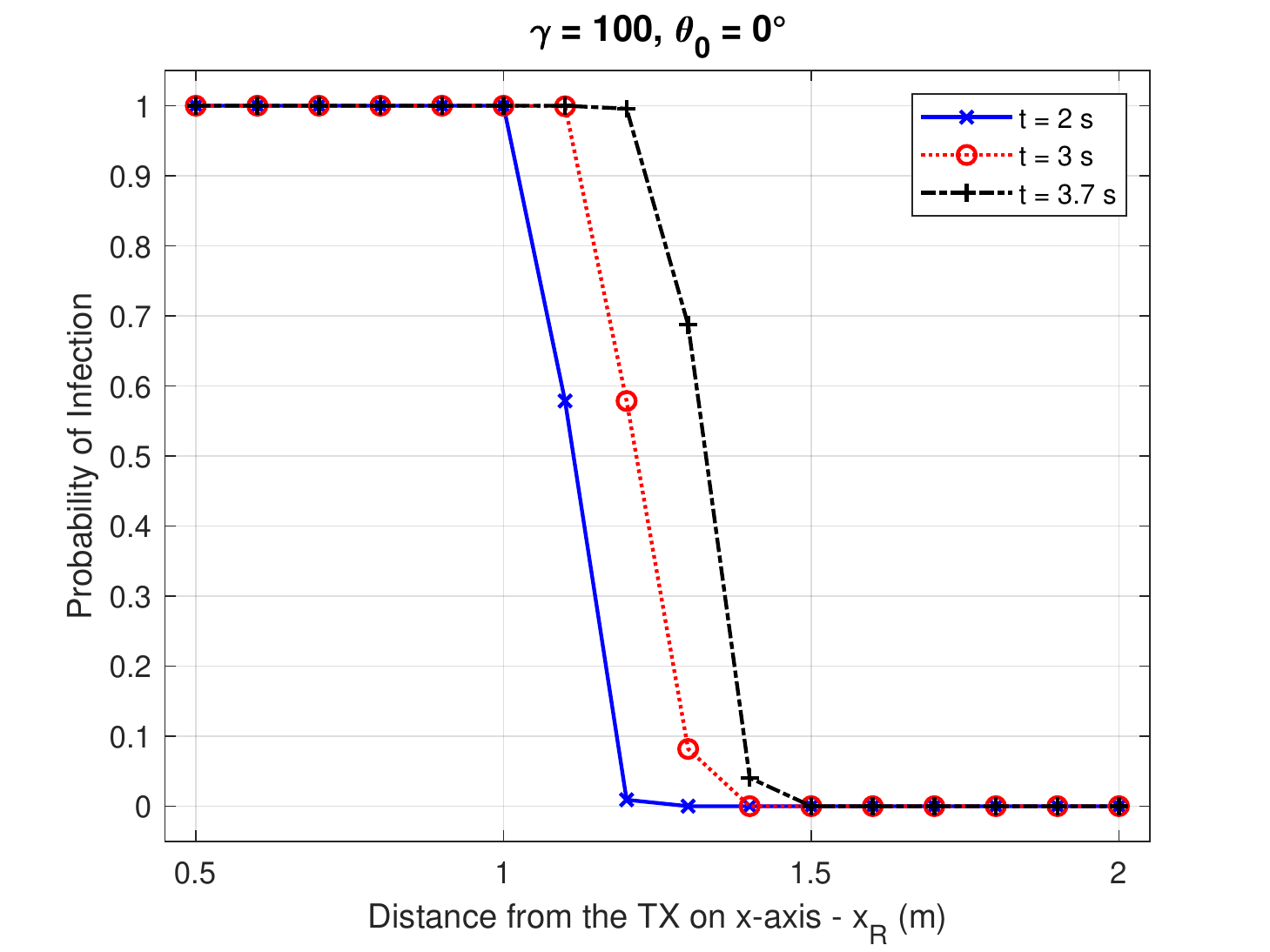}}
	\caption{Probability of infection according to distance for different time values.}
	\label{P_inf}
	\vspace{-0.7cm}
\end{figure}

\begin{table}[!b]
		\vspace{-0.5cm}
	\centering
	\caption{Experimental parameters}
	\scalebox{0.85}{
	\begin{tabular}{p{45pt}|p{81pt}|p{43pt}|p{74pt}}
	\hline
	\textbf{Parameter}	& \textbf{Value} & \textbf{Parameter}	& \textbf{Value}\\
	\hline  \hline 
	$\Delta t$  &  $ 0.1$ s & $t_s$ & $10$ s \\
	TX's position & ($0$,$1.7$,$0$) m ($x$,$y$,$z$)& $v_{c_{(1,..K),0}}$ & $11.2$ m/s (cough) \cite{zhu2006study}\\
 	RX's position & ($1.7$,$0$) m ($y$,$z$)& $g$ & $9.81$ m/s$^2$  \\
 	$ I_0 $ & $ 0.0131$ kg m/s \cite{bourouiba2014violent} & $\alpha_e$ & $ 0.2116 $ \cite{tang2009schlieren} \\
 	$ F_0 $ & $ 0.0023$ kg m/s$^2$ \cite{bourouiba2014violent} & $\beta_{bb}$ (female) & $8.853$ cm \cite{young1993head} \\
 	$ \rho_d $ & $993$ kg/m$^3$ \cite{nicas2005toward}& $\beta_{bb}$ (male) & $9.131$ cm  \cite{young1993head} \\
 	$\rho_f$ (at $34^\circ C$) & $0.98$ kg/m$^3$ \cite{picard2008revised} & $\beta_{ss}$ (female) &$6.901$ cm \cite{young1993head}\\
 	$\rho_a$ (at $23^\circ C$) & $1.172$ kg/m$^3$ \cite{bourouiba2014violent}  & $\beta_{ss}$ (male) & $7.57$ cm \cite{young1993head}\\
 	$\mu_a$ & $19\hspace{-0.07cm} \times \hspace{-0.07cm} 10^{-6}$ kg/(m s) \hspace{-0.1cm} \cite{bourouiba2014violent} & & \\ 	
	\hline   \hline           
	\end{tabular}}
	\label{Sim_parameters}
\end{table} 

\begin{table}[!bhp]
		\vspace{-0.5cm}
	\centering
	\caption{Initial number of droplets \cite{duguid1946size}}
	\scalebox{0.85}{
		\begin{tabular}{>{\raggedright}p{25pt}|p{35pt}||>{\raggedright}p{25pt}|p{35pt}||>{\raggedright}p{25pt}|p{35pt}}
			\hline
			\textbf{$\mathbf{d}$ ($\mu$m)}	& \textbf{Quantity (cough)} &  \textbf{$\mathbf{d}$ ($\mu$m)}	& \textbf{Quantity (cough)} & \textbf{$\mathbf{d}$ ($\mu$m)}	& \textbf{Quantity (cough)}\\
			\hline  \hline 
			$2$ & $50$ & $40$ & $240$ & $200$ & $35$ \\
			$4$	& $290$ & $50$ & $110$ & $250$ & $29$ \\
			$8$ & $970$ & $75$ & $140$ & $500$ & $34$ \\
			$16$ & $1600$& $100$ & $85$ & $1000$ & $12$ \\
			$24$ & $870$ & $125$ & $48$ & $2000$ & $2$ \\
			$32$ & $420$ & $150$ & $38$ & Total & 4973\\			
			\hline   \hline           
	\end{tabular}}
	\label{Droplet_sizes}
\end{table}

For $x_R \hspace{-0.1cm}=\hspace{-0.1cm} 1.5$ m and $\theta_0 \hspace{-0.1cm} = \hspace{-0.1cm} 0^\circ$, Figs. \ref{Traj_R} (a) and (b) depict the trajectories of the cloud with respect to RX in 3-D and 2-D, respectively. Due to initial momentum of coughing and net buoyant force, the cloud propagates in the horizontal and vertical directions, respectively. As also observed in these trajectories, the 3-D cloud can encompass (or intersect with) the RX and the RX is exposed to droplets during the passage of the cloud. Fig. \ref{Traj_R} (c) shows the interaction of a human who can be infected with a disease-spreading human by quantifying the number of droplets received from the cloud. The number of droplets are given by their mean and their variations (vertical bars) due to the Gaussian distribution. These variations are calculated as three times the standard deviation ($99.73\%$ confidence interval) for each sample. As observed in Fig. \ref{Traj_R} (c), the number of droplets in cloud decreases, since large-sized droplets settle due to the gravity. As mentioned in Section \ref{EESM}, the RX gets infected, when the received number of droplets is above $ \gamma $ as plotted in Fig. \ref{Traj_R} (c). This figure shows the importance of the exposure time which is the interval of the changing zone in the received signal. If the RX is exposed to the cloud less, it is possible not to be infected, since the received number of droplets can be below the threshold. This interaction is also clarified in Fig. \ref{P_inf} by showing the relation of the infection probability with the distance of the RX to the TX for different propagation time instances. Actually, this figure reveals that if the RX is exposed to the cloud for a longer period, the RX is more likely to be infected. 

Different scenarios can be analyzed by employing the infection state, which is the output of the end-to-end system model, for various $x_R$, $\gamma$, $\theta_0$ values and male/female receivers. In Fig. \ref{Inf_s} (a), the threshold is set to zero to determine the safe zone where there is no possibility of infection. This safe zone starts at $x_R=1.7$ m which also shows the minimum social distance. Fig. \ref{Inf_s} (c) shows that initial coughing angle affects the infection state severely. For $\theta_0$ values between $0$ and $-25$ degrees, it is more likely to infect someone due to the buoyant forces affecting the propagation of the cough cloud. Therefore, it is safer to cough with an initial angle $\theta_0 \leq -25^\circ$ which is depicted as safe coughing angle in Fig. \ref{Inf_s} (c). Coughing with $\theta_0 \geq 0^\circ$ may not be safe, since small droplets (aerosols) can suspend in the air and settle eventually due to gravity or drift due to the indoor air currents in the long term. In Figs. \ref{Inf_s} (a) and (c), the results are indistinguishable for male and female receivers. However, Fig. \ref{Inf_s} (b) shows that the infection states of female and male humans can be affected differently for the same $\gamma$ values. Actually, this figure reveals that female humans are less likely to get infected, even if their immune system's strength are the same with male humans due to the slight difference in the face dimensions.
\vspace{-0.25cm}
\section{Conclusion}\label{Conc}
In this paper, an algorithmic end-to-end system model is proposed for droplet-based communication via coughing/sneezing between two static humans for an indoor scenario. The TX emits a cloud which is a mixture of droplets and air and it propagates under the influence of the initial momentum, gravity and buoyancy. A receiver model which defines the central part of the human face as the RX cross-section is proposed for the reception of droplets to give an output of infection state of the RX. The  transmitted and received number of droplets are modeled as random processes which lead us to derive the probability of infection. Numerical results show that the safe zone for the RX starts at $1.7$ m for a horizontally coughing TX. It is also revealed that the initial coughing angle of the TX, the distance between the TX and RX and the detection threshold which actually corresponds to the strength of the human immune system are significant parameters to model the airborne pathogen transmission. Furthermore, the reception of pathogens can be affected by the sex of the human. As the future work, we plan to extend our study for channel parameter estimation and modeling end-to-end communication for mobile TXs and RXs.

\vspace{-0.25cm}
\appendix
For a settling droplet, the downwards net force for a spherical droplet at the $i^{th}$ time step with the diameter $d_k$ is given as
\begin{equation}
F_{down_i} = G_i - B_i = V_d \rho_d g - V_d \rho_a g =  \frac{\pi d_{k,i}^3}{6} (\rho_d - \rho_a) g \label{F_d1}
\end{equation}
Furthermore, an upward drag force acts in the opposite direction of gravity due to the interaction of the droplet with the air. This upward drag force is given as \cite{munson2009fundamentals}
\begin{equation}
F_{up_i} = \frac{1}{2} \rho_d v_s^2 \pi \frac{d_{k,i}^2}{4} C_D
\end{equation}
where $C_D$ is the drag coefficient. For the settling condition, these upward and downward forces are in equilibrium. Hence, we can obtain the settling velocity by equating these two forces and pulling out $v_s$ as given by
\begin{equation}
v_{s_{k,i}} = \sqrt{\frac{4 d_{k,i} g (\rho_d - \rho_a)}{3 \rho_d C_D}}.\label{v_ss}
\end{equation}
Here, $C_D$ changes according to $Re$ as given by
\cite{reuter2005metrics}
\begin{numcases}
{\hspace{-0.6cm}C_D =} 
\frac{24}{Re} &\hspace{-0.7cm}$, Re<2$ \hspace{0cm} \label{C_D1}
\textrm{(Stokes flow)} \\ 
\frac{18.5}{Re^{3/5}} &\hspace{-0.7cm}$, 2 \leq Re \leq 500$ \label{C_D2}  \hspace{0cm} \textrm{(Intermediate flow)}  \\ 
0.44 &\hspace{-0.7cm}, $500 < Re \leq 2\times 10^5$ \label{C_D3} \hspace{-0.3cm} \textrm{(Newton's flow)}
\end{numcases}
When the drag coefficients in (\ref{C_D1})-(\ref{C_D3}) and $Re$ in (\ref{Re}) are substituted into (\ref{v_ss}), the settling velocities in (\ref{v_s1})-(\ref{v_s3}) can be obtained.

\bibliographystyle{ieeetran}
\bibliography{ref_fg_dbc}

\vspace{-0.5cm}
\begin{IEEEbiography}[{\includegraphics[width=1.08in,height=1.95in,clip,keepaspectratio]{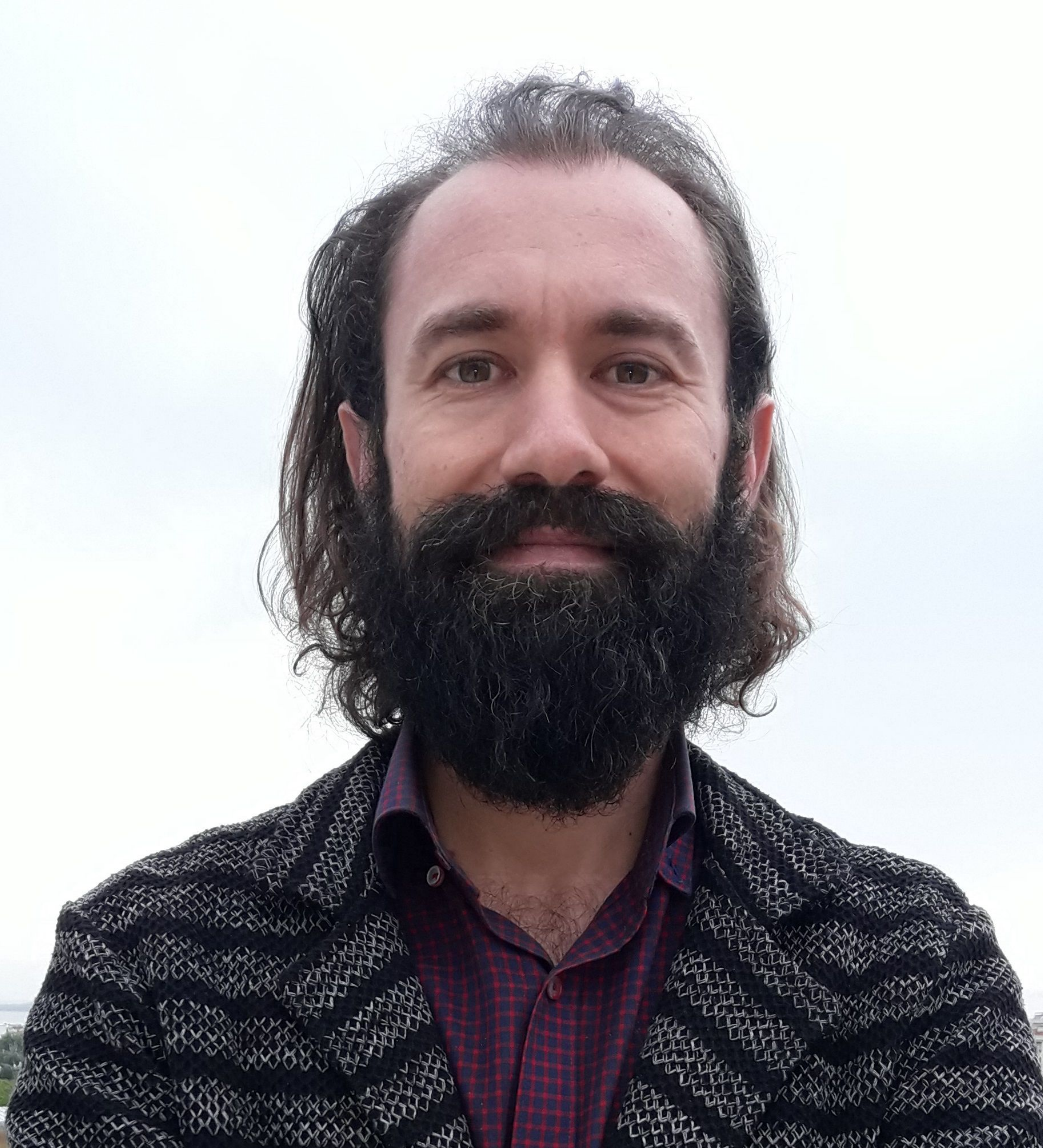}}]{Fatih Gulec}
received his B.Sc. and M.Sc. degree from Gazi University, Ankara, Turkey in 2007 and 2015, respectively both in electrical and electronics engineering. He is currently pursuing the Ph.D. degree in İzmir Institute of Technology, İzmir, Turkey as a research/teaching assistant under the supervision of Assoc. Prof. Dr. Barış Atakan. His research interests include micro and macroscale molecular communications and molecular networks. 
\end{IEEEbiography}
\vspace{-1cm}
\begin{IEEEbiography}
[{\includegraphics[width=1.08in,height=1.4in,clip,keepaspectratio]{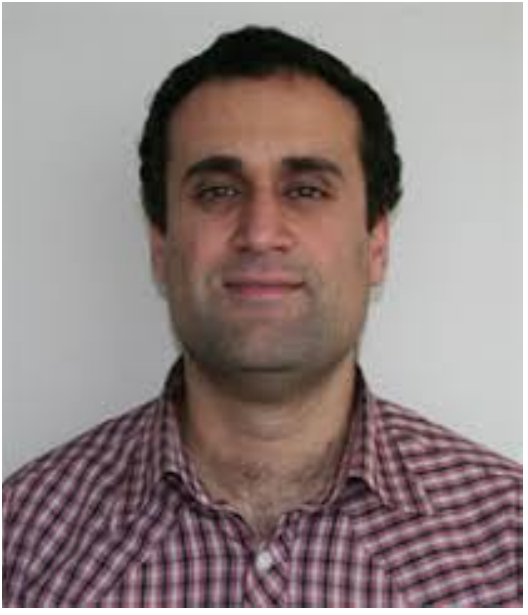}}]{Baris Atakan} received the B.Sc. degree from Ankara University, Ankara, Turkey, in 2000, the M.Sc.  degree from Middle East Technical University, Ankara, in 2005, and the Ph.D. degree from the Next-Generation and Wireless Communications Laboratory, School of Sciences and Engineering, Koç University, Istanbul, Turkey, in 2011, all in electrical and electronics engineering. He is currently an Associate Professor with the Department of Electrical and Electronics Engineering, İzmir Institute of Technology, İzmir, Turkey. His current research	interests include nanoscale and molecular communications, nanonetworks and biologically inspired communications.
\end{IEEEbiography}




\end{document}